\documentclass[12pt,preprint]{aastex}

\newcommand {\kms} {\,{\rm km\,s}^{-1}}

\usepackage{rotating}

\shorttitle{BSSs in NGC~6397}
\shortauthors{Lovisi et al.}
 
\begin{document} 
%\title{Spectroscopic analysis of the Blue Straggler population in NGC~6397}

\title{Chemical and kinematical properties of BSSs and HB stars in NGC 6397
\footnote{Based on FLAMES observations collected at the European Southern Observatory, proposal numbers 
073.D-0058, 075.D-0125 and 081.D-0356.}}
\author{
L. Lovisi\altaffilmark{1},
A. Mucciarelli\altaffilmark{1}, 
B. Lanzoni\altaffilmark{1},
F.R. Ferraro\altaffilmark{1}, 
R. Gratton\altaffilmark{2},
E. Dalessandro\altaffilmark{1},
R. Contreras Ramos\altaffilmark{1}}
\affil{\altaffilmark{1} Dipartimento di Astronomia, Universit\`a degli Studi
di Bologna, via Ranzani 1, I--40127 Bologna, Italy}

\affil{\altaffilmark{2} INAF- Osservatorio Astronomico di Padova,
  Vicolo dell'Osservatorio 5, I--35122 Padova, Italy}
\date{18 January 2012}
 
\begin{abstract} 
We used three sets of high-resolution spectra acquired with the
multifiber facility FLAMES at the Very Large Telescope of the European
Southern Observatory to investigate the chemical and kinematical
properties of a sample of 42 horizontal branch (HB) stars, 18 Blue
Straggler Stars (BSSs) and 86 main sequence turn-off and sub-giant
branch stars in the nearby globular cluster NGC 6397.  We measured
rotational velocities and Fe, O and Mg abundances. All the unevolved stars
in our sample turn out to have low rotational velocites ($v \sin i< 10\kms$), 
while HB stars and BSSs show a broad distribution, with values ranging from 0 to $\sim
70 \kms$. For HB stars with $T<10500$ K there is a clear 
temperature-oxygen anti-correlation, that can be understood if
the star position along the HB is mainly determined by the He content.
The hottest BSSs and HB stars (with temperatures $T>8200$ K
and $T> 10500$ K, respectively) also show significant deviations in
their iron abundance with respect to the cluster metallicity (as traced by
the unevolved stars, [Fe/H]=-2.12).  While similar chemical
patterns have been already observed in other hot HB stars, this is the
first evidence ever collected for BSSs.  We interprete these abundance
anomalies as due to the metal radiative levitation, occurring in stars
with shallow or no convective envelopes.
\end{abstract} 
 

\section{Introduction}
\label{intro}
Star clusters have been long considered to host coeval and chemically
homogeneous stars \citep{Renzini1986}. The traditional paradigm of a
striking chemical homogeneity in globular cluster (GC) stars is still
valid in terms of the iron content, with only two notable exceptions
known to date: $\omega$Centauri \citep{NorrisDaCosta1995,
Johnson2010} and Terzan 5 \citep{Ferraro2009a, Origlia2011}, which are now
thought to be the remnants of giant systems instead of genuine GCs
\footnote{Significantly smaller dispersion (0.1-0.2 dex) in the iron content
has been measured in M54 \citep{Carretta2010a} 
M22 \citep{Marino2009, Marino2011a} and NGC~1851 
\citep{Carretta2010b, Carretta2011}.}.
Instead, large star-to-star abundance variations in the light elements
(such as C, N, O, Na, Mg, Al; see \citealt{Gratton2004} for a review)
and strong anti-correlations (for example, between Na and O, or
between Mg and Al) have been revealed in the last decade or so by
high-resolution spectroscopic studies of main sequence (MS) and red
giant stars in a large sample of GCs both in the Milky Way
\citep{Carretta2010c} and in the Magellanic Clouds
\citep{Mucciarelli2009}. The scenario generally invoked to explain these
inhomogeneities is based on a second generation of stars formed during
the first few 100 Myr of the cluster life, from intra-cluster medium
polluted by the first stellar population. Intermediate-mass
asymptotic giant branch stars and MS fast rotating massive stars have
been proposed as the most likely polluters.

Large abundance anomalies have also been observed on the horizontal
branch (HB) of some GCs: HB stars cooler than $\sim$11000 K show abundances
consistent with those of giants \citep{Glaspey86, Glaspey89,
  Lambert92, Cohen97, Behr1999, Behr2000a, Peterson2000} whereas for
HB stars hotter than $\sim$11000 K, departures from the overall cluster
abundances are found \citep{Glaspey89, Behr1999,Behr2000a,
  Peterson1995, Peterson2000, Moehler2000, Fabbian2005, Pace2006}. 
In particular, a deficiency of He and an overabundance (up to solar and
super-solar values) of elements heavier than Ne have been observed
and they have been interpreted in terms of the gravitational settling
of He and the radiative levitation of heavy elements (see
\citealt{Michaud1983}), competing with mixing processes (like
rotationally induced meridional circulation, mass loss and turbulent
diffusion) able to moderate the final effects. 
Moreover, the discontinuity in chemical abundances leads to a
discontinuity in Str\"{o}mgren u magnitudes and 
(u-y) colors (the so-called ``Grundahl jump'', see  
\citealt{Grundhal1999}), where the more metallic HB stars 
appear to be brighter than the predictions of canonical ZAHB models.
Analogous mechanisms are also invoked to explain the origin of the peculiar chemical
patterns observed in Population I stars: MS stars of A and F spectral
type in some open clusters (OCs) \citep{VarenneMonier1999, Gebran2010,
  Gebran2008, GebranMonier2008} and in the Ursa Major group
\citep{Monier2005} exhibit chemical abundance anomalies similar to
those observed in GC HB stars.  In particular, enhancement of iron
peak elements has been observed, and large star-to-star scatter has
been measured for many other elements, like C, O, Na, Mn, Ca, Sc, Sr
and Ba.

Chemical anomalies can also be expected for Blue Straggler Stars
(BSSs).  These objects are brighter and bluer (hotter) than the MS
turn-off (TO), they lie along an extension of the MS in the
color-magnitude diagram (CMD), and they are thought to be generated by
the coalescence of two MS stars via two main channels: mass transfer
(MT) activity in a binary system \citep{mcrea64_mtbss1,
  zinn76_mtbss2}, or stellar collisions (COL) and merging
\citep{hills76_colbss}. BSSs generated by both channels can coexist
within the same cluster \citep{Ferraro2009b}. Anomalies in the surface chemical abundances
are predicted for MT-BSSs \citep{Lombardi1995, SarnaDeGreve1996}, since
the accreted material should come from the inner region of the donor
star where nuclear processing has occured and, in particular, C and O
have been depleted.  No significant mixing between the inner core and
the outer envelope is expected, instead, in the case of COL-BSSs.
Indeed, C and O depletion has been observed in a sub-sample of BSSs in
47~Tucanae (\citealt{Ferraro2006}, hereafter F06), consistently with
the predictions of the MT scenario.

Accurate high-resolution spectroscopic studies not only revealed the
presence of all the chemical anomalies discussed above, but also
showed that different stellar populations in GCs and OCs can behave
differently also in terms of kinematical properties, in particular in
terms of rotation.  In fact, while TO and sub-giant branch stars in
GCs always show low rotational velocities (few $\kms$ only;
\citealt{LucatelloGratton2003}), a bimodal rotational velocity
distribution has been observed for HB stars: the cooler HB stars
(T$<$11000 K) have rotations between 10 and 40 $\kms$ (much faster
than old MS stars), while hotter HB stars show a
markedly different behaviour and rotate at less than 10 $\kms$
(\citealt{Peterson1995}; \citealt{Behr2000a}; \citealt{Behr2000b};
\citealt{RecioBlanco2002, RecioBlanco2004}). 
% The high rotation ofcool HB objects could be explained by assuming that the progenitor red
% giant branch stars had rapidly rotating cores and differential rotation in their
% convective envelopes, and that angular momentum is redistributed from
% the former to the latter \citep{SillsPinsonneault2000}. Above a
% critical value of the equatorial rotational velocity, the meridional
% circulation could prevent the gravitational settling of He and the
% radiative levitation of Fe-peak elements, thus explaining the absence
% of chemical anomalies in HB stars with T$<$11000 K (see above). The
% decrease in rotation rates toward higher temperatures is not predicted
% by the models, but could be interpreted as a result of the
% gravitational settling, which creates a mean molecular
% weight gradient that inhibits angular momentum transport in the star
% \citep{Vauclair1999}. 
Even if the picture is not still completely clear, there are hints that
rotational rates and gravitational settling are strictly related and possibly self-powered. 
In fact, the high rotation of cool HB objects could be explained by assuming that the progenitor red
giant branch stars had rapidly rotating cores and differential rotation in their
convective envelopes, and that angular momentum is redistributed from
the former to the latter \citep{SillsPinsonneault2000}. Moreover, above a
critical value of the equatorial rotational velocity, the meridional
circulation could prevent the gravitational settling of He and the
radiative levitation of Fe-peak elements (which are already inefficient for these stars), 
thus contributing to explain the lack of chemical anomalies in HB stars with T$<$11000 K (see above). 
On the contrary, the decrease in rotation rates for HB stars toward higher temperatures is not predicted
by the models. \citet{Vauclair1999} suggests that it could be interpreted as a result of the
gravitational settling, creating a mean molecular weight gradient that inhibits the angular momentum 
transport in the star. An alternative scenario to explain the decreasing rotation rates toward higher 
temperature for the HB stars is provided by \citet{VinkCassisi}: according to their theory,
the radiative levitation of the heavy elements sets a stellar wind that can significantly
remove angular momentum.
Rotational velocities of A and F stars in OCs
are typically lower than 100-150 $\kms$, and, at odds with HB stars,
no trend with stellar temperatures or chemical abundances is observed
\citep{GebranMonier2008, Gebran2008, Gebran2010}. Such a behaviour
well agrees with the predictions of some models
\citep{Charbonneau1991} where the timescales of diffusion processes
(including radiative levitation) are much shorter than those of
rotational mixing.

Concerning BSSs, a rapid rotation is expected in the MT scenario
because of the angular momentum transfer
\citep{SarnaDeGreve1996}. High rotational velocities are also expected
for COL-BSSs \citep{BenzHills1987}, but some magnetic braking or disk
locking mechanisms might intervene to significantly decrease the
rotation \citep{LeonardLivio1995, Sills2005}. Unfortunately, the
efficiencies and timescales of these mechanisms are still unknown,
thus preventing a clear prediction of the expected rotational
velocities. From an observational point of view, only one BSS rotating
at $\simeq$ 100 $\kms$ and no correlation between CO-depletion and
rotational values have been found in 47 Tucanae (F06). Instead,
the largest sample of fast rotating BSSs ever observed in a GC has been
revealed in M4 by \citet{Lovisi2010}, (hereafter L10; $\sim 40\%$ of the 
surveyed BSSs have rotational velocities larger than 50 $\kms$).

In this framework we started an extensive study of the chemical and
the kinematical properties of TO, HB  stars and BSSs in GCs by using the
Very Large Telescope (VLT) of the European Southern Observatory
(ESO). The present paper reports on the results obtained in NGC 6397 
and it is organized as follows. The observations
are described in Sect. \ref{obs}.  Data reduction procedures and the
determination of stellar radial velocities and cluster membership are
discussed in Sect. \ref{red}.  The attribution of atmospherical
parameters to all targets is described in Sect. \ref{par}. The
measured rotational velocities are presented in Sect. \ref{rot}.
Sect. \ref{chem_an} and \ref{abu} describe, respectively, methods
and results of the chemical abundance analysis for our
sample. Finally our conclusions are drawn in Sect. \ref{discuss}.

\section{Observations}
\label{obs}
This work is based on the analysis of spectra of individual stars in
NGC~6397, obtained with the multi-object facility FLAMES
mounted at the ESO VLT. Observations were performed
during two different runs (073.D-0058(A) and 081.D-0356(A), hereafter
P73 and P81, respectively) with the UVES+GIRAFFE combined mode, during 
three nights between April and July 2004 for P73 and three nights in June 2008 for P81.
Spectra have been acquired for 33 BSSs and 42 HB
stars, most of them being observed in both P73 and P81. In order to
compare the results obtained for BSSs and HB stars with a sample of
unevolved cluster stars, we also analysed archive GIRAFFE data
obtained within program 075.D-0125(A) (hereafter P75), including
spectra for 86 turn-off and sub-giant branch stars (hereafter both
called TO stars), already discussed by \citet{Lind2008}. The GIRAFFE
gratings and exposure times used in the various observing runs are
listed in Table \ref{gratings}.  During P73 we also observed six BSSs with the UVES Red
Arm 580nm, that provides a wavelength coverage from 4800 to 6800
\AA\ with a resolution of $R\simeq40 000$.

\begin{figure}
\plotone{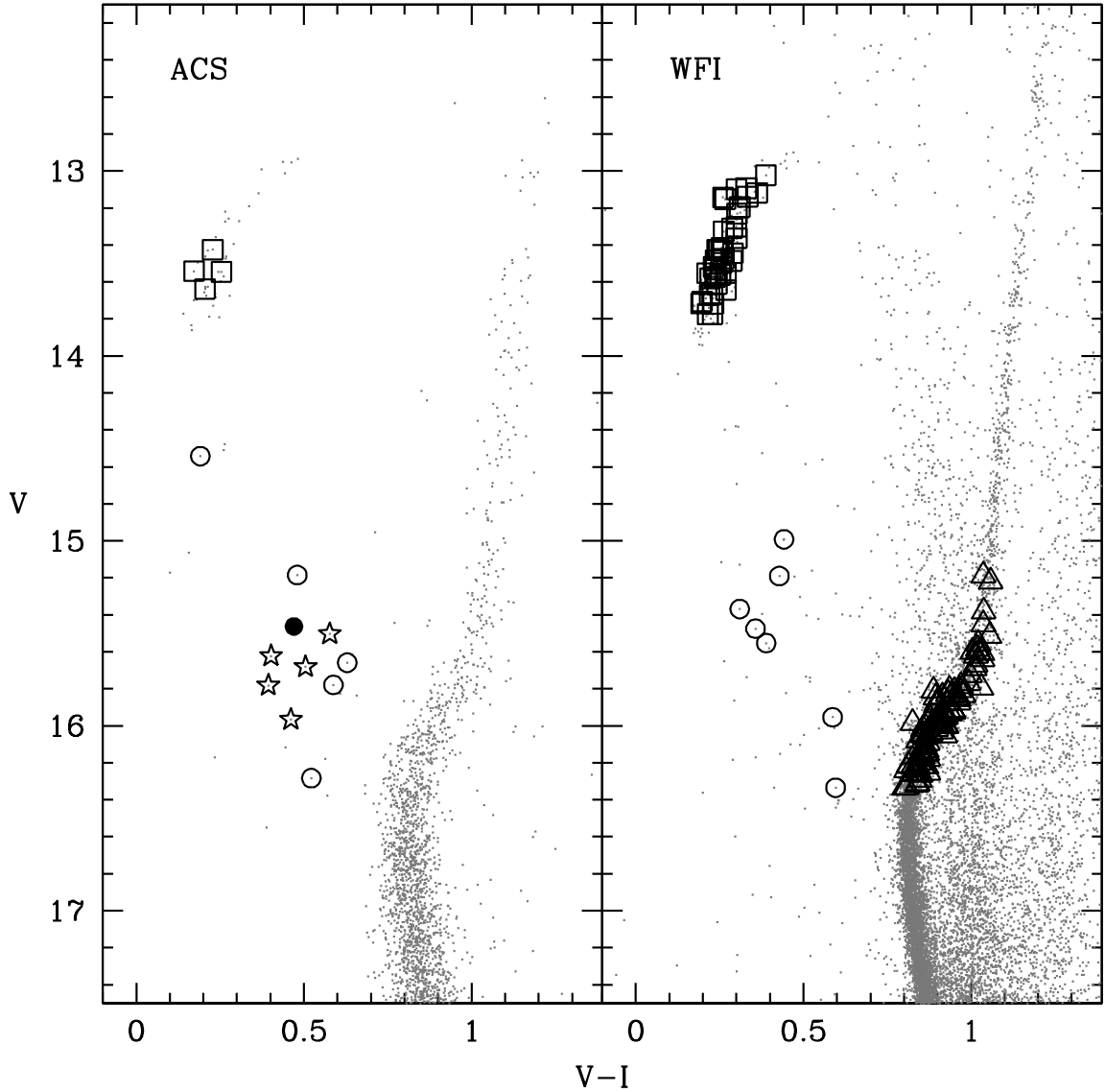}
\caption{Spectroscopic targets in the CMDs of NGC~6397 for the ACS and WFI datasets.
Open triangles and squares are TO and HB respectively, whereas large circles and stars are BSSs. 
In particular, star symbols mark the BSSs that show radial velocity variations
and are also identified as SX Phe by \citet{Kaluzny2003}. The filled black 
circle represents the fast rotating BSS \#1100126 (that is also a SX Phe).} 
\label{cmd}
\end{figure}

The spectroscopic target selection has been performed from a photometric catalog obtained by 
combining ACS@HST data for the central region (within
$r < 140''$ from the cluster center) and WFI@ESO observations for the outer
region (Contreras Ramos et al. 2012, in preparation).
In order to avoid contamination from spurious light in the spectrograph fibers,
only the most isolated stars have been selected: we conservatively
excluded targets having stellar sources of comparable or brighter
luminosity within 3 arcsec. Fig. \ref{cmd} shows the CMDs 
for our ACS and WFI datasets, with the position of all the analysed targets.

\section{Data Reduction, radial velocities and cluster membership}
\label{red}
All the spectra acquired during P73 and P81 have been reduced by using the last version of the GIRAFFE 
and UVES ESO pipelines \footnote{http://www.eso.org/sci/software/pipelines/}, whereas for the P75 data 
pre-reduced spectra have been retrieved from the GIRAFFE archive maintained at the 
Paris Observatory \footnote{http://giraffe-archive.obspm.fr/}. The reduction procedure includes
the bias subtraction, flat-field correction, computation of the dispersion solution by using a Th-Ar reference 
lamp and finally the extraction of the one-dimensional spectra.
When it was possible, the accuracy of the wavelength calibration has been checked by
measuring the position of some  telluric emission lines \citep{Osterbrock1996}. 
The master sky spectrum, needed to sample the sky contribution, was computed by averaging several 
spectra of sky regions without bright stellar sources
and was subtracted from each exposure. 

In order to measure the radial velocities (RVs) of our targets,
we have used the IRAF task \textit{fxcor}
and some tens of Fe~I and Fe~II lines, the Mg~II line at $\lambda \approx$ 4481 \AA\  
and the O~I triplet at $\lambda \approx$ 7774 \AA\ . The H$\alpha$ Balmer line was used 
for stars that do not show significant lines in other gratings (because of the very high temperatures,
high rotational velocities and$/$or low signal to noise ratio). The task \textit{fxcor} is based on 
the Fourier cross-correlation between the target spectra 
and a template of known radial shift \citep{TonryDavis1979}. As templates for the sample of BSSs,
HB and TO stars we used synthetic spectra computed with the photometric parameters of each target
(see details in Sect. \ref{chem_an}).

\begin{figure}
\plotone{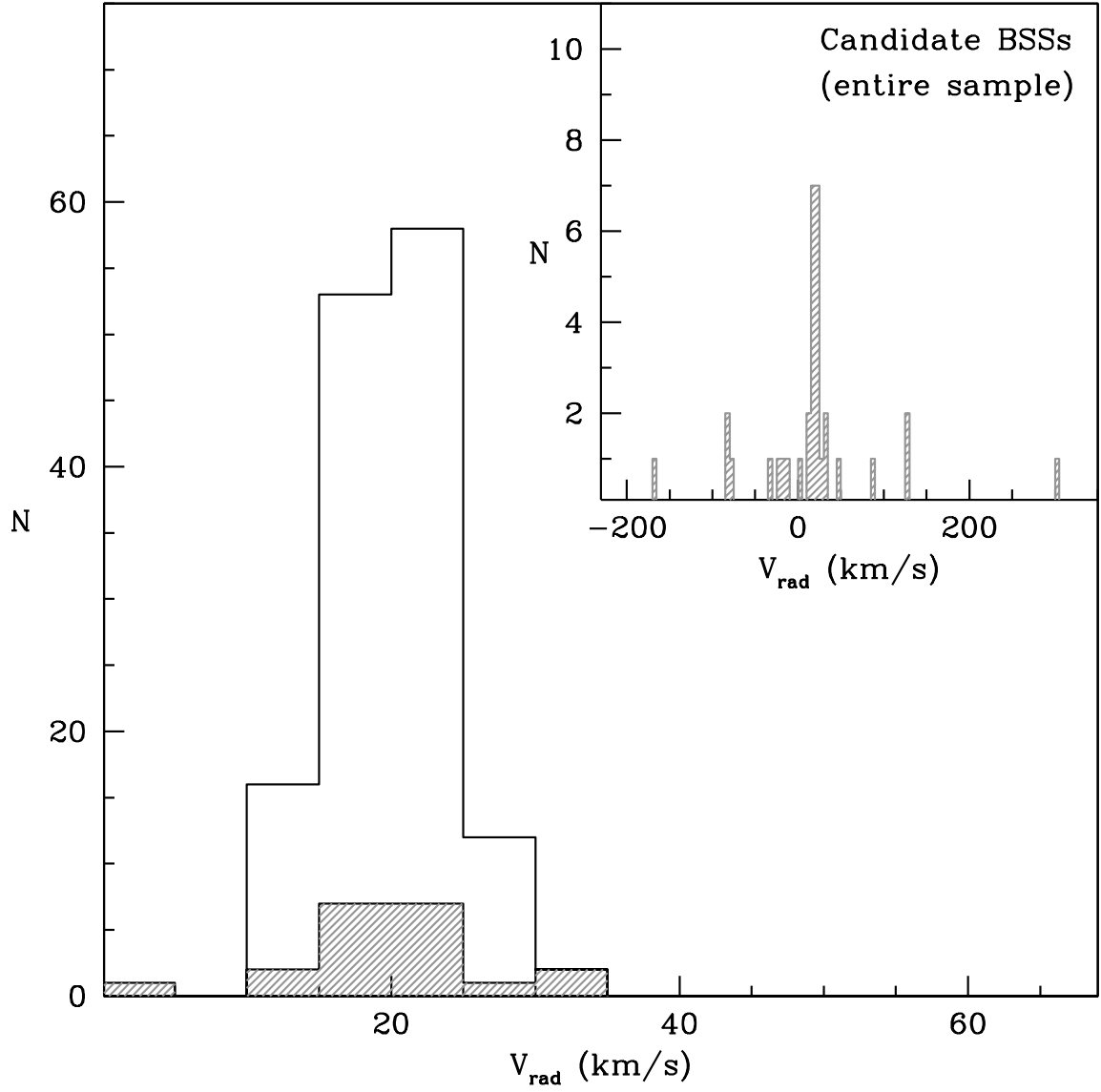}
\caption{Radial velocity distribution for the TO+HB and BSSs samples.
The inset shows the RV distribution for all the BSSs. The empty histogram in the main panel represents the RV distribution
for the TO+HB stars, whereas the shade-histogram represents the BSS distribution after the $\sigma$-rejection procedure 
discussed in the text.}
\label{vrad}
\end{figure}

The RV distribution of the TO+HB stars is shown in Fig. \ref{vrad}.
The mean value is $\langle RV\rangle = 20.3\pm0.3\ (\sigma=3.8)\ \kms$,
in good agreement with previous studies (see e.g. \citealt{Harris1996, Milone2006, 
Lind2008, Hubrig2009, GonzalezHernandez2010}). 
This has been adopted as the systemic velocity of NGC~6397 and used to assign the cluster membership to each star:
only stars having radial velocity in agreement with this value within 3$\sigma$
have been considered as members of NGC~6397.

The inset of Fig. \ref{vrad} shows the radial velocity distribution for the entire sample of BSSs.
As evident, BSSs span a wide range of values 
from $\sim$-200 $\kms$ up to $\sim$+300 $\kms$, with the bulk of population between 0 $\kms$ and 30 $\kms$.
The application of the $\sigma$-rejection algorithm with respect to the cluster systemic velocity
yields to exclude 15 objects, which probably are field stars. 
For all of them, the RV resulting from each exposure exceed the systemic velocity by more than 3$\sigma$. The remaining
18 BSSs display the RV distribution shown (as shaded histogram) in the main panel of Fig. \ref{vrad},
with a mean value $\langle RV\rangle = 20.1\pm1.5\ (\sigma=6.9)\ \kms$.
Only two (namely \#81828 and \#1100063) out of six BSSs observed with UVES fibers result to be cluster members.
We also found that BSSs \#2200239, \#1100127, \#110126 and \#1100170 correspond to stars
V10, V11, V15 and V23, respectively, in \citet{Kaluzny2003} and are classified as variable SX Phe stars.
For all the BSSs classified as cluster members,
we checked the RV values (from each exposure) as a function of the Julian Day, to search for any 
relevant variability of RV. We found that only 4 BSSs (\#1100126, \#1100162, \#1100208, \#2200239)
exhibit small RV variations. Furthermore, all RV variable stars are grouped in a restricted region
of the CMD, close to the location of SX Phe stars. It is then probable that all these stars are pulsating
variables. RV values together with other relevant information for the BSS, HB and TO samples are listed in Table \ref{bss},
Table \ref{hb} and Table \ref{to} respectively.

\begin{figure}
\plotone{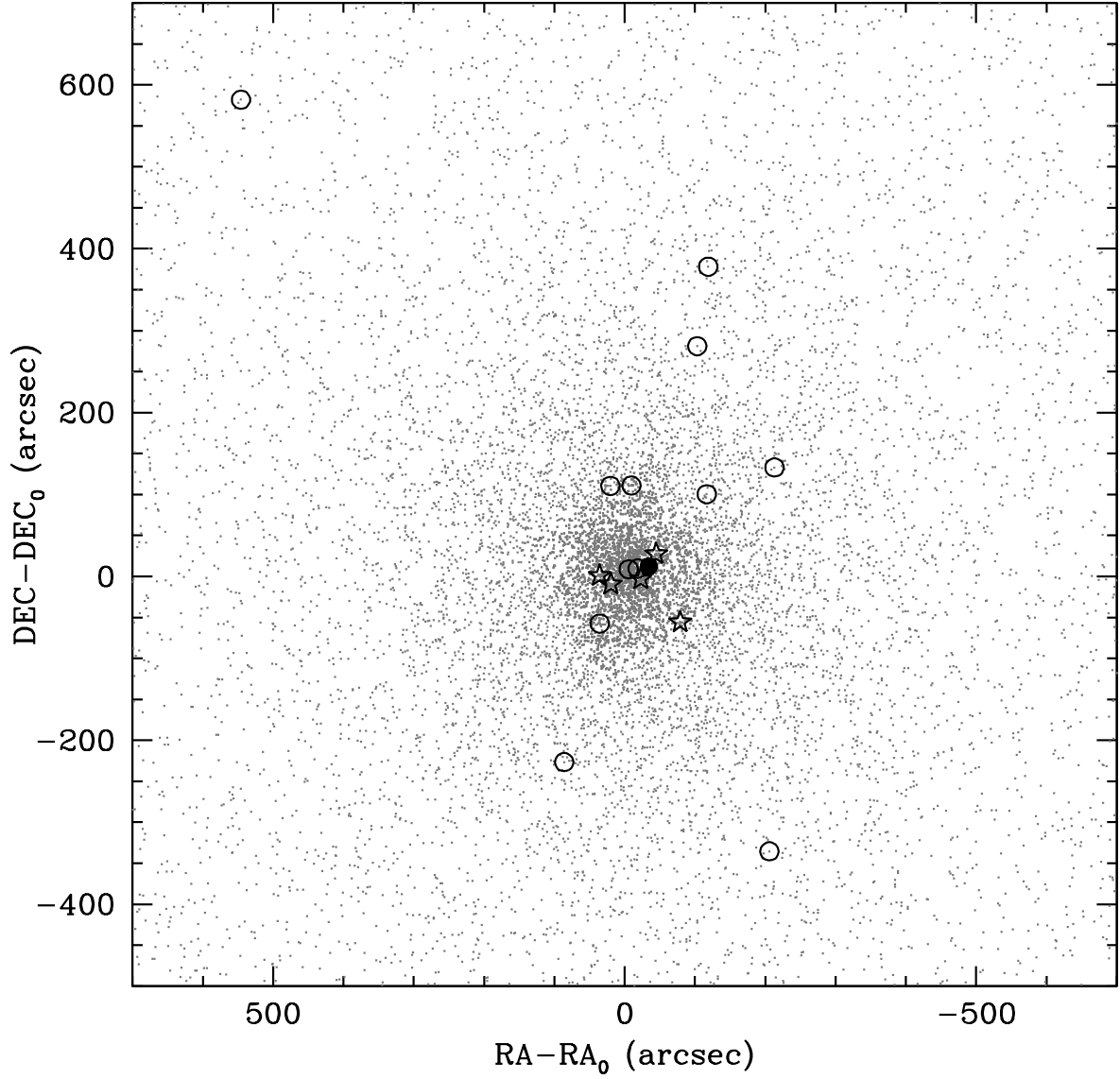}
\caption{Position of member BSSs with respect to the cluster center (symbols have the same meaning of Fig. \ref{cmd}).}
\label{map}
\end{figure}

For all the member stars we combined the rest-frame spectra of each exposure thus obtaining a medium spectrum
with signal-to-noise ratio S/N $\simeq$ 50-100 for BSSs,
S/N $\simeq$ 150-200 for HB stars and S/N $\simeq$ 50-70 for TO stars.
Finally, Fig. \ref{map} shows the position with respect to the cluster center for all member BSSs.

\section{Atmospherical parameters}
\label{par}

Atmospherical parameters for our targets have been derived photometrically,
according to their position in the CMD and the comparison with theoretical stellar models 
taken from the BaSTI database \citep{Pietrinferni2006}.
The best-fit isochrone of the cluster sequences (with age=13.5 Gyr, metallicity Z=0.0003, 
$\alpha$-enhanced chemical mixture and assuming a distance modulus 
of 11.92 and E(B-V)=0.19, in agreement with \citealt{Ferraro1999}) 
has been used to infer temperatures 
(T) and gravities ($\log g$) for TO stars (see Table \ref{to}). 
The ZAHB model taken from the same database has been used to infer parameters and masses of the HB stars (see Table \ref{hb}).
Finally, by projecting the star position on a set of isochrones with different ages,
we derived T, $\log g$ and masses for the BSSs sample (see Table \ref{bss}).
In previous papers (see for example L10) we derived effective temperatures for BSSs
by fitting the wings of the H$\alpha$ Balmer line.
However this technique is more reliable in the range
T$\sim$ 5500-8000 K, where the broadening of the wings is driven by the
self-resonance broadening of the H atoms. For higher temperatures the H$\alpha$ wings mainly suffer
from the Stark broadening and become 
less sensitive to temperature variations.
According to the photometric determination, the majority of the BSSs in our sample are
hotter than 8000 K, their temperatures ranging from $\sim$ 7400 K up to $\sim$ 13000 K
that is in the range where the sensitivity of the H$\alpha$ is reduced.
% However this technique is applicable only in the range
% T$\sim$ 5500-8000 K, where the broadening of the wings is driven by the
% self-resonance broadening of the H atoms. For higher temperatures the Balmer wings mainly suffer
% from the Stark broadening and become
%basically insensitive to T. 
%According to the photometric determination, the majority of the BSSs in our sample are
%hotter than 8000 K, their temperatures ranging from $\sim$ 7400 K up to $\sim$ 13000 K,
%so that T estimate through the Balmer wings is not recommended.

Errors in temperatures and gravities were computed by assuming typical uncertainties
in magnitude and colors for all the targets
(we conservatively estimate $\delta$V $\lesssim$0.05 mag and $\delta$(V-I)$\lesssim$0.07 mag
for TO and BSSs and $\delta$V $\lesssim$0.02 mag and $\delta$(V-I)$\lesssim$0.028 mag for HB stars).
Typical values for temperature errors are 50-150 K for BSSs,
 100 K for TO and 50-80 K for HB.
The uncertainty on the $\log g$ determination is substantially negligible
(lower than 0.1) for all the targets. 

The microturbulent velocity is generally derived spectroscopically, by requiring that weak and strong lines 
of a given species (usually Fe) provide the same abundance within the uncertainties.
In most of our targets, however, the number of available Fe lines is not large enough 
to permit a direct determination of this parameter.
% We therefore assumed 0 $\kms$ for the BSSs and HB stars,
% according to the values derived by \citet{Pace2006} and \citet{Hubrig2009} for 
%  HB stars of similar spectral type and in agreement with the fact 
% that hot stars (with spectral type A or earlier) have radiative atmospheres, 
% with no convection or large-scale flows. 
% The analysis of the two BSSs observed with UVES,
% for which we have a large number of Fe lines of different intensity, fully confirms
% that the adopted value is well appropriate for the hot stars in our sample.
We therefore used the relation by \citet{ForSneden2010} to assign a microturbulent 
velocity value to our HB stars according to their temperature. For 
the BSS sample, we assumed 0 $\kms$ for all the targets. In fact, the analysis of the two BSSs observed with UVES
(that have temperatures of 8299 K and 13183 K, covering almost the entire temperature range of the BSS sample)
for which a large number of Fe lines with different strength are available, fully confirms
that the adopted value is well appropriate for the BSSs.
Finally, we assumed 1.5 $\kms$ as microturbulent velocity for TO stars (according to L10).
A conservative error of 1 $\kms$ has been assumed for the microturbulent velocity
of BSSs and HB, and 0.5 $\kms$ for TO stars.

\section{Rotational velocities} 
\label{rot}

\begin{figure}
\plotone{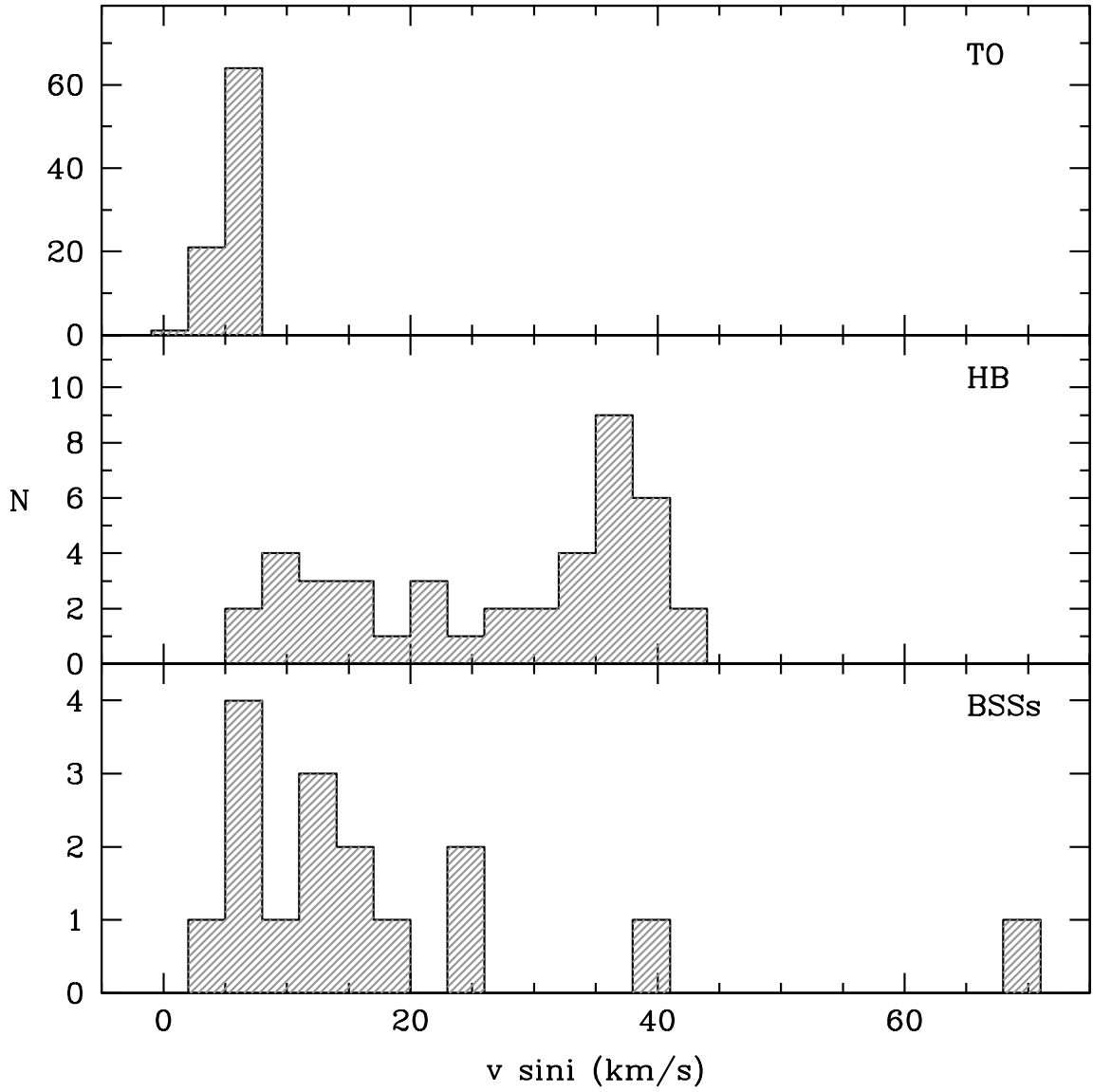}
\caption{Rotational velocity distribution for TO, HB stars and BSSs.}
\label{istobi}
\end{figure}

Projected rotational velocities ($v \sin i$) were measured from the analysis of the most
prominent atomic lines (namely the Mg~II line and the O~I triplet in the GIRAFFE spectra and
 some Fe~I line in the UVES ones)
\footnote{Only for two stars for which the only observed spectral feature is the H$\alpha$,
we could not measure $v \sin i$. In fact, the line profile is insensitive to rotational 
velocities smaller than 100 $\kms$.}.
We performed a $\chi^{2}$ minimization between the observed spectrum and a
grid of synthetic spectra, computed with different values of rotational velocities
and by taking into account
the instrumental profile, the microturbulent and 
the macroturbulent velocity and the Doppler broadening. The instrumental 
profile has been derived by measuring the FWHM of bright unsaturated lines 
in the reference Th-Ar calibration lamp \citep[see i.e.][]{Behr2000a}.
The macroturbulent velocity was assumed to be 0 $\kms$ for BSSs and HB stars,
whereas for TO stars the formula provided by
\citet{Gray1984} has been used. Finally, the Doppler and the microturbulent velocity broadenings
have been computed by using the atmospherical parameters discussed in Sect. \ref{par}.
In order to test the accuracy of our method we performed 
Monte Carlo simulations of 100 synthetic spectra with $v\sin i$ =
0, 5, 10, 15 and 20 $\kms$, with the typical resolution and 
S/N of our observations. We recovered the original values within less than 2 $\kms$. 
Uncertainties on rotational velocities are 3-4 $\kms$ for TO, 1-2 $\kms$ for HB and 1 $\kms$ for the bulk of BSSs.
The values derived for 16 BSSs, HB and TO stars are listed in Table \ref{bss}, \ref{hb} and \ref{to}, respectively. The
distributions obtained for the different samples are compared in Fig. \ref{istobi}.
All the TO stars display very low rotational velocities, peaked at $\sim7\kms$ and never 
exceeding $\sim10\kms$ (in agreement with the values derived by \citealt{LucatelloGratton2003}).
In comparison, the HB sample exhibits a much broader distribution: only few HB stars show rotational velocity
around $\sim 10 \kms$ and $v \sin i$ can be as high as $\sim42\kms$. The only 
previous analysis for HB stars in NGC~6397 is provided 
by \citet{Hubrig2009} that measured $v \sin i\sim8-10\kms$ for three blue HB stars (T$>$11500 K).\\
The rotational velocities distribution for BSSs also results to be very broad, ranging from $\sim0\kms$ up to 
$70 \kms$, with an average value of $\langle v \sin i\rangle$=18.2$\pm1.0 \kms$ ($\sigma$=16.7) and a median value of 13 $\kms$. 
In particular, BSS \#1100126 (which is shown in Fig. \ref{cmd} as a filled circle)
is the most fast rotating star in our sample, with $v \sin i=70\kms$.
Interestingly enough, among BSSs with T$<$8500 K RV variable stars (included SX Phe) show the largest
rotational velocities, whereas the other ones have values compatible with the TO distribution.
While no trend between rotational velocity and temperature is observed for TO and HB stars,
a systematic increase is found for BSSs (Fig. \ref{levconfgrey}, top panel). The two hottest
BSSs, however, show low rotation (less than 20 $\kms$).

\section{Chemical analysis}
\label{chem_an}

Iron, magnesium and oxygen abundances have been derived for almost all our targets (according to the available gratings):
several tens of Fe~I and Fe~II lines, the Mg~II line at 4481 \AA\ and the O~I triplet 
at $\sim$ 7774 \AA\ have been used. Chemical abundances have been derived by using both the measured equivalent widths (EWs)
and spectral synthesis through the comparison with synthetic spectra.

In order to calculate abundances for all the elements, the set of codes developed by R.L.Kurucz \citep{Kurucz1993, Sbordone2004} has been used:
model atmospheres have been computed by using ATLAS9, whereas WIDTH9 and SYNTHE have been used to obtain 
chemical abundances from the measured EWs and to compute synthetic spectra, respectively.
In particular, ATLAS9 model atmospheres have been computed under the assumption of LTE plane-parallel geometry and
by adopting the new opacity distribution functions by \citet{CastelliKurucz2003}, without the 
inclusion of the {\sl approximate overshooting} (see \citealt{Castelli1997}).
Atomic data for all the lines are from the most updated version of the Kurucz line list by F.Castelli
\footnote{http://wwwuser.oat.ts.astro.it/castelli/linelists.html}. 
Finally, the used reference solar abundances are from \citet{GrevesseSauval1998}
for iron and magnesium, and from \citet{Caffau2011} for the oxygen.

The EWs of Fe~I, Fe~II and O~I were measured with our own 
code that fits the absorption lines with a Gaussian profile (see L10 and \citealt{Mucciarelli2011});
the abundance calculation was then performed with the code WITDH9.
For the Mg~II line (an unresolved blending of three close components belonging to the same 
multiplet) we derived abundances through a $\chi^{2}$ minimization between the observed spectra and a grid of synthetic 
spectra.
Only for BSS \#1100126 (the most fast rotating star of the sample) all the abundances were derived 
by comparison with synthetic spectra, because the observed line profile significantly deviates from the 
Gaussian approximation.

An important point to recall (at least for the BSSs and HB stars analysed here) is the possible departure
from the LTE assumption: in fact, the photospheric layers of hot stars are exposed to the strong UV radiation 
coming from the stellar interior and this effect is magnified in metal-poor stars because of the low opacity of their 
photospheres. The magnitude and the sign of the non-LTE corrections are still highly uncertain, due to the 
incompleteness of the used atom models (in particular for iron) 
and the uncertainties about the collisions rate with the H~I atoms. 
For the computation of the O abundances we included non-LTE corrections taken from the statistical equilibrium calculations of 
\citet{Takeda1997}. Instead, for the Mg and Fe abundances, 
no grid of non-LTE corrections are available in the literature for the range of parameters typical of our targets. 
\citet{Mashonkina2011} computed LTE and non-LTE abundances for some iron transitions 
in A and F spectral type stars, pointing out that non-LTE corrections are relevant for Fe~I lines, whereas Fe~II 
lines are negligibly affected by these effects (at the level of a few hundredths of dex).
Thus, in the following we will use the abundances derived from Fe~II lines as proxies of the iron content of BSSs and HB stars, 
in order to minimize the non-LTE effects. On the contrary, the iron content of TO stars is derived from 
Fe~I lines, because these transitions are much more numerous with respect to the Fe~II ones and because the 
non-LTE corrections for these lines should be negligible in dwarf stars (see e.g. \citealt{Gratton1999}).
Concerning the magnesium, \citet{Przybilla2001} discussed the line formation for this feature, pointing out 
that large ($>$0.2-0.3 dex) corrections are expected at least for A type stars with solar metallicity. 
Since however no corrections are available for the Mg~II line at 4481 \AA\ , Mg abundances have 
not been corrected for non-LTE effects.

For BSS \#1100063 (observed with UVES) we have been able to measure also the He~I line at 5876 \AA\ , 
a transition that is visible only in stars hotter than $\sim$9000 K. 
The He abundance has been derived through spectral synthesis. To generate the grid of synthetic spectra 
with different He abundances, we have computed for this star a number of suitable model atmospheres by using the 
ATLAS12 code \citep{Castelli2005} that allows to calculate model atmospheres with arbitrary 
chemical compositions through the use of the {\sl opacity sampling} method, at variance with ATLAS9.

\section{Chemical abundances}
\label{abu}
 
{\bf Iron --}
For the TO stars we find an average iron abundance [Fe/H]=-2.12$\pm$0.01 ($\sigma$=0.08) dex, 
in very good agreement with the values published in the literature \citep{Harris1996, Thevenin2001, 
Gratton2001, Lind2008}.
However, for BSSs and HB stars we obtain larger values and significantly higher dispersions.
In fact, for the 11 BSSs for which we measured Fe~II lines, we obtain [Fe/H] = -1.20 $\pm$  0.22 ($\sigma$ = 0.74) dex
and for the entire HB sample we find [Fe/H]= -1.94 $\pm$ 0.14 ($\sigma$ = 0.63) dex.
The dispersions observed in the BSS and HB samples are completely incompatible with the
uncertainties of the measures. 
In particular, as shown in Fig. \ref{levconfgrey}, the iron abundances of BSSs exhibit a systematic trend with the 
temperature: the coldest BSSs (with T$\sim$7400-8200 K) have [Fe/H]=-1.88 $\pm$ 0.14 dex, that 
is marginally in agreement with the TO iron content, whereas for T$>$8200 K, [Fe/H] increases 
up to solar values. A milder but similar behaviour is observed also along the HB sequence: stars with T$<$10500 K 
have an average iron content of [Fe/H]=-2.16$\pm$0.02 dex ($\sigma$ = 0.10), whereas two of the three stars with 
T $>$10500 K have an approximately solar iron abundance. No trend between [Fe/H] and rotational velocity is 
found for BSSs (see Fig. \ref{rotfeall}).

{\bf Magnesium --}
A similar behaviour is also found for the BSSs magnesium abundances, whereas no trend between [Mg/H]
and T is detected for HB and TO stars (see Fig. \ref{mgabb}, top panel). In terms of
[Mg/Fe] (lower panel in Fig. \ref{mgabb}) these evidences translate in a mild decrease of this parameter for increasing BSS
and HB temperature, and a constancy of it for the TO population, with an average value of
[Mg/Fe]$\sim$0.2 dex. 
This value is $\sim$0.2 dex lower than that provided by \citet{Carretta2009a}.
However, we remind that our Mg abundances do not include corrections for non-LTE, that 
could explain such a discrepancy.

{\bf Oxygen --}
HB stars show a large dispersion and a clear trend between
O abundances and temperature (Fig. \ref{oxabb}), with [O/Fe] decreasing from enhanced 
values (0.75 dex) to sub-solar values (-0.47 dex) for increasing temperature.
Almost all outliers in this temperature-oxygen anti-correlation are over luminous
stars, likely evolved HB stars. Such a distribution highlights
the intrinsic star-to-star scatter observed in all the GCs where light 
elements have been studied so far (see e.g. \citealt{Carretta2009b}) that is usually interpreted 
in the framework of self-enrichment processes occuring in the early stages of GC formation. 
We note that the range of the derived [O/Fe] ratios well matches with the distribution 
recently discussed by \citet{lind2011} that find [O/Fe]=+0.71 dex for the first generation stars and 
[O/Fe]=+0.56 dex for the second generation stars.
The trend with effective temperature (corresponding to the star mass along the HB) 
is analogous to that observed in M4 \citep{Marino2011b}, NGC~2808 \citep{Gratton2011}
and NGC~1851 \citep{Gratton2012}, even if the scatter is higher. Note that low [O/Fe] abundance ratio observed 
in some of the hottest HB stars could be partially due to the radiative levitation effects that 
take place. 
It may be interpreted as due to an anti-correlation between O and He abundances, taking into account
the expected anti-correlation between the mass of evolving stars and the He
content (see \citealt{Dantona2004}). For BSSs we detect a mild decrease of [O/H]
with T and one star (namely \#79976) shows a very high oxygen abundance 
([O/H]=-0.44 dex), higher than those measured for the other BSSs and the HB stars. 
Its spectrum is shown in Fig. \ref{oxrich}, top panel. 

{\bf Helium --} 
BSS \#1100063 exhibits a He~I line weaker than that predicted for a standard He abundance.
In fact, the He~I line (broadened by a moderate rotational velocity) is well reproduced with a helium mass fraction 
of Y=0.001, whereas it is totally inconsistent with Y=0.25 (Fig. \ref{oxrich}, bottom panel). 

\section{Discussion}
\label{discuss}
The kinematical properties and chemical abundances here derived for
the TO sample are in good agreement with previous
determinations. They indicate that TO stars do not
significantly rotate ($v \sin i\sim 7 \kms$) and are chemically
homogeneous (at least in the iron and magnesium content), 
with very small dispersions around average values
of [Fe/H]=-2.12 dex and [Mg/Fe]= 0.17 dex (see Sect. \ref{abu}). 

On the other hand, the hot populations exhibit significantly different
properties, in terms of both the rotational velocity and the chemical
abundances. 
% The rotational distribution for the HB stars in our sample 
% is very wide and possibly shows a main peak at $\sim 15-20\kms$ and a second peak at 
% $\sim 40\kms$, similar to what found by \citet{Peterson1995} and \cite{Behr2000a}.
% Unfortunately, there are no HB stars hotter than 11000 K in our sample
% so that we are not able to observe the bimodality in the rotational
% distribution already found in other GCs. 
The rotational distribution for the HB stars in our sample 
is very wide and shows a spread distribution for values lower than $\sim 25-30\kms$ and a peak at 
$\sim 35-40\kms$, similar to that found by \citet{Peterson1995} and \cite{Behr2000a}.
Unfortunately, there are no HB stars hotter than 11000 K in our sample
so that we are not able to observe the bimodality in the rotational
distribution already found in other GCs. 

A wide rotational distribution is also observed for the BSS
population, with values of $v \sin i$ ranging from 0 to $70\kms$.
Although the statistics is low (only 16 stars in total), 
it is interesting to note that the two hottest stars
in the sample (T $>$ 9200 K) show rotational velocities lower than 20 $\kms$
at odds with BSSs cooler than 9200 K which show a wide rotation distribution 
with a possible trend with temperature (the rotational velocity increasing with 
temperature). Additionally, no trend is observed between BSS rotational velocities 
and the [Fe/H] ratio (see Fig. \ref{rotfeall}).

The distribution of $v \sin i$ obtained for the BSS population in NGC
6397 is qualitatively similar to that found in the other GCs studied
so far, namely 47 Tuc and M4 \citep[see Fig. 3 in F06, and Fig. 2 in
  L10, respectively; for 47 Tuc also see][]{DeMarco2005, Shara1997}. In
fact, in these GCs the rotation distribution is peaked at low values
(consistent with the TO star velocities) and shows a long tail toward
larger values. Quantitatively, however, significant differences can be
recognized, especially in terms of the shape of the distribution, the
highest measured values and the percentage of fast rotators. As shown in 
Fig. \ref{rotconf} in fact,
the distribution of $v \sin i$ (between 0 and 20-30 $\kms$) is clearly peaked 
in 47 Tuc and M4, while in NGC 6397 it is more evenly spread. In addition,
while values as high as $100-150\kms$ are found for a few BSSs in the
two other clusters, the maximum measured value in NGC 6397 is $v \sin
i=70\kms$. According to L10, by defining as ``fast rotators'' the BSSs spinning faster
than $50\kms$ , only 6-7\% of such stars are found in NGC 6397 and
47 Tuc, while this fraction raises to 40\% in M4.  Clearly, the
interpretation of such findings is not straightforward and more
statistics is needed to explain these differences. The interpretation
is unclear also in terms of the BSS formation channels, since
conflicting predictions are still provided by the available
theoretical models. In fact, high rotation rates are expected, at
least in the early evolutionary stages, for BSSs formed by either
direct collisions or mass transfer activity in binary systems. 
However, precise and solid theoretical predictions are
still lacking, and possibly some (poorly constrained) braking mechanisms likely
start to play a role in slowing down these stars during their
subsequent evolution.  \citet{LeonardLivio1995} predict that a
convective zone able to slow down the star develops in the envelope of
collisional BSSs, while the same phenomenon does not occurr in the
models of \citet{Sills2005}. The latter, instead, predict that a
magnetically locked accretion disk forms around the star and is able
to remove at least a fraction of the stellar angular momentum.
Indeed, high rotational velocities could be just a transient
phenomenon, characterizing only the early stages of (some) BSS life;
then the rotation could slow down until those stars become
indistinguishable from the BSSs generated with low rotational
velocities. Such a scenario could partially explain the different
fraction of fast rotating BSSs observed in 47 Tuc, M4 and NGC 6397.

\begin{figure}
\plotone{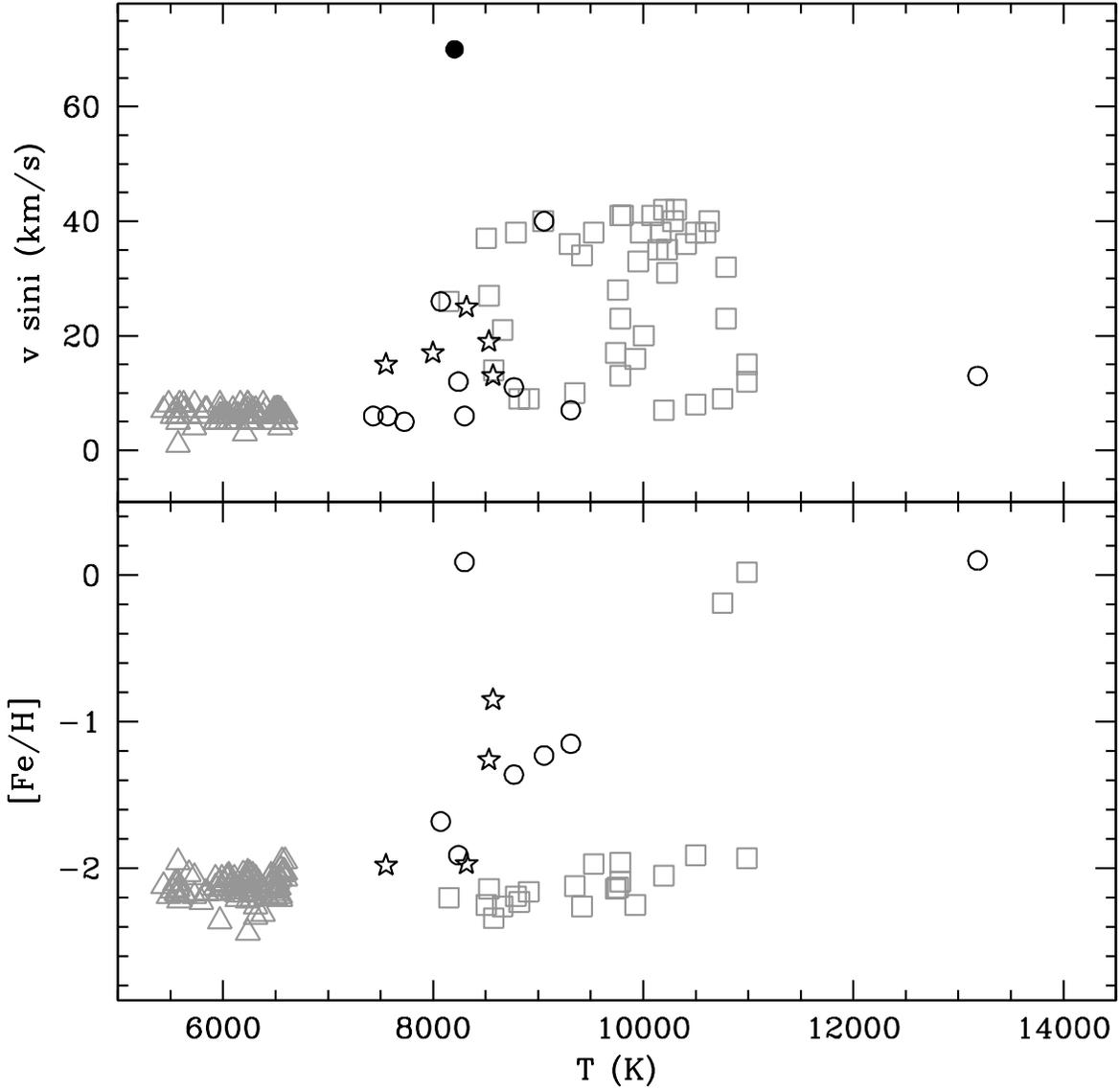}
\caption{Rotational velocities (top panel) and [Fe/H] (bottom panel) as a function of the stellar temperatures for TO 
(empty grey triangles), HB (empty grey squares) and BSSs (same symbols used in the previous figures). Values and errors
for all the targets are listed in tables \ref{bss}, \ref{hb} and \ref{to}.}
\label{levconfgrey}
\end{figure}

\begin{figure}
\plotone{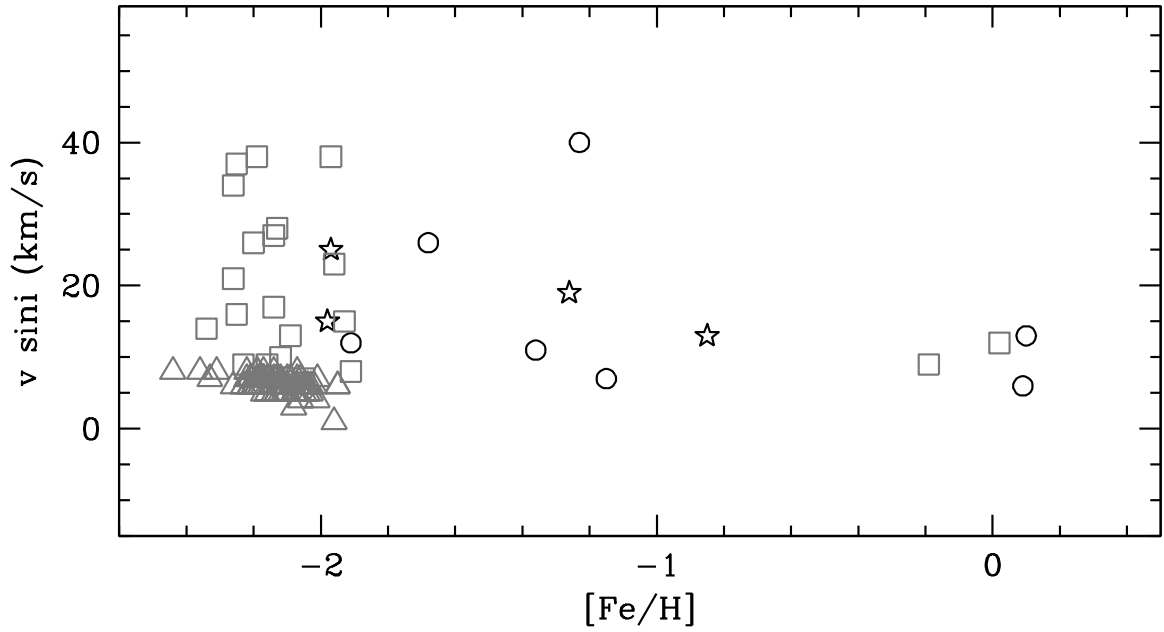}
\caption{Rotational velocities as a function of the [Fe/H] for TO, HB stars and BSSs. Different symbols
have the same meaning of Fig. \ref{cmd}.}
\label{rotfeall}
\end{figure}

\begin{figure}
\plotone{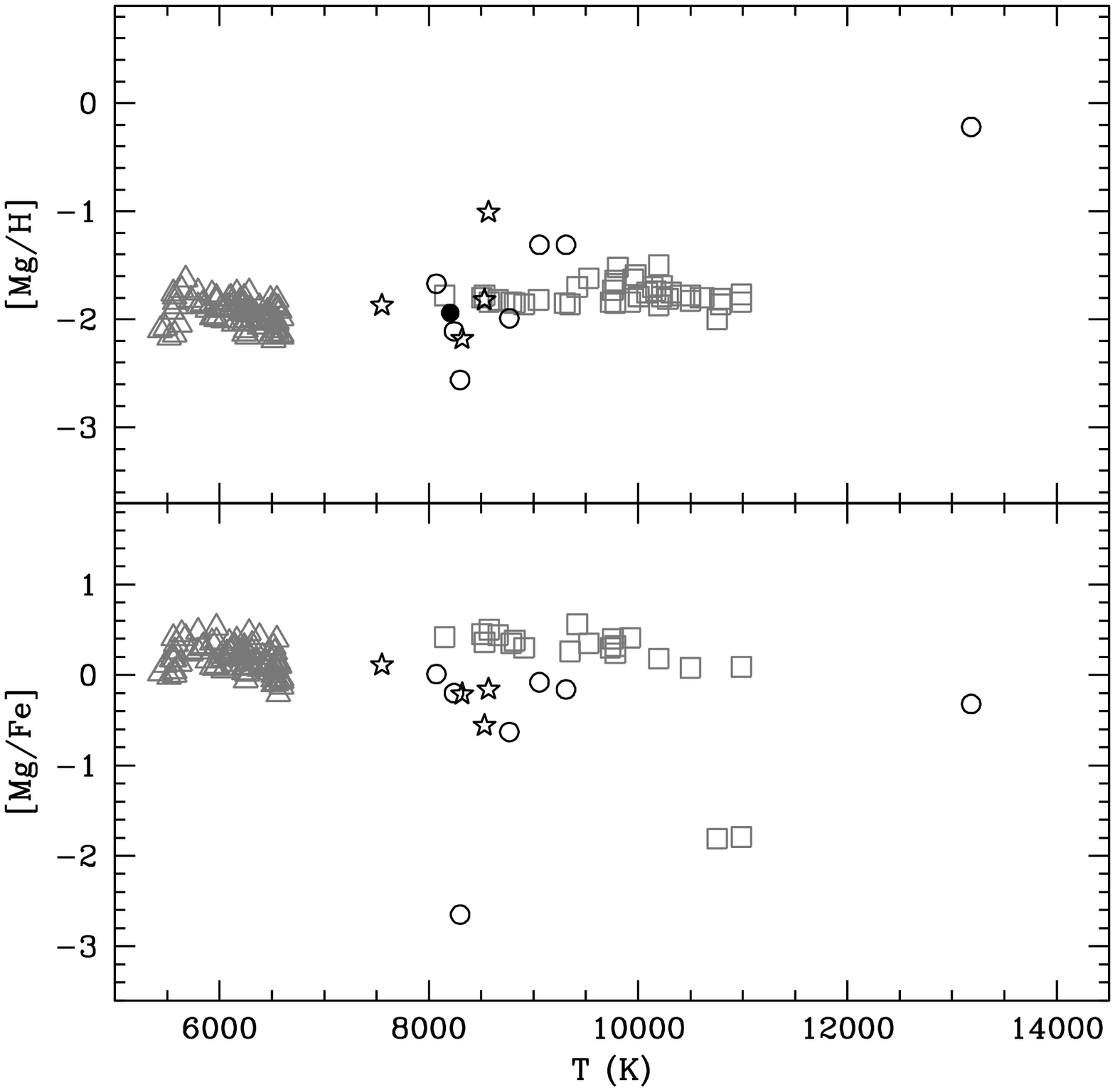}
\caption{[Mg/H] and [Mg/Fe] as a function of stellar temperatures for TO, HB stars and BSSs (marked with the same symbols used in previous figures).
Value and errors for BSS, HB and TO stars are listed in tables \ref{bss}, \ref{hb} and \ref{to}.}
\label{mgabb}
\end{figure}

\begin{figure}
\plotone{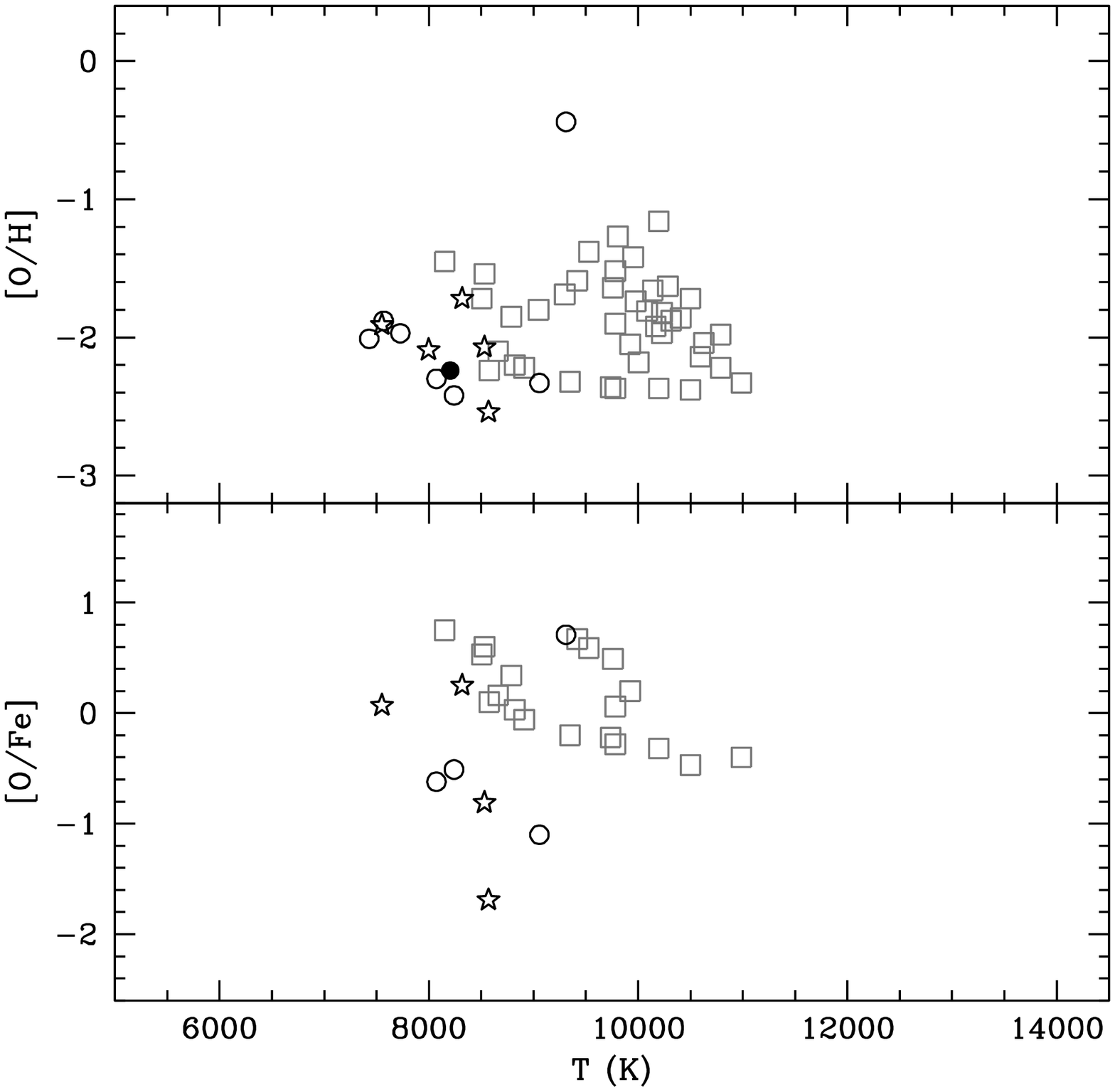}
\caption{[O/H] and [O/Fe] as a function of the stellar temperatures for HB stars and BSSs (marked with the same symbols used in previous figures).
Value and errors for BSS and HB stars are listed in tables \ref{bss} and \ref{hb}.}
\label{oxabb}
\end{figure}

\begin{figure}
\plotone{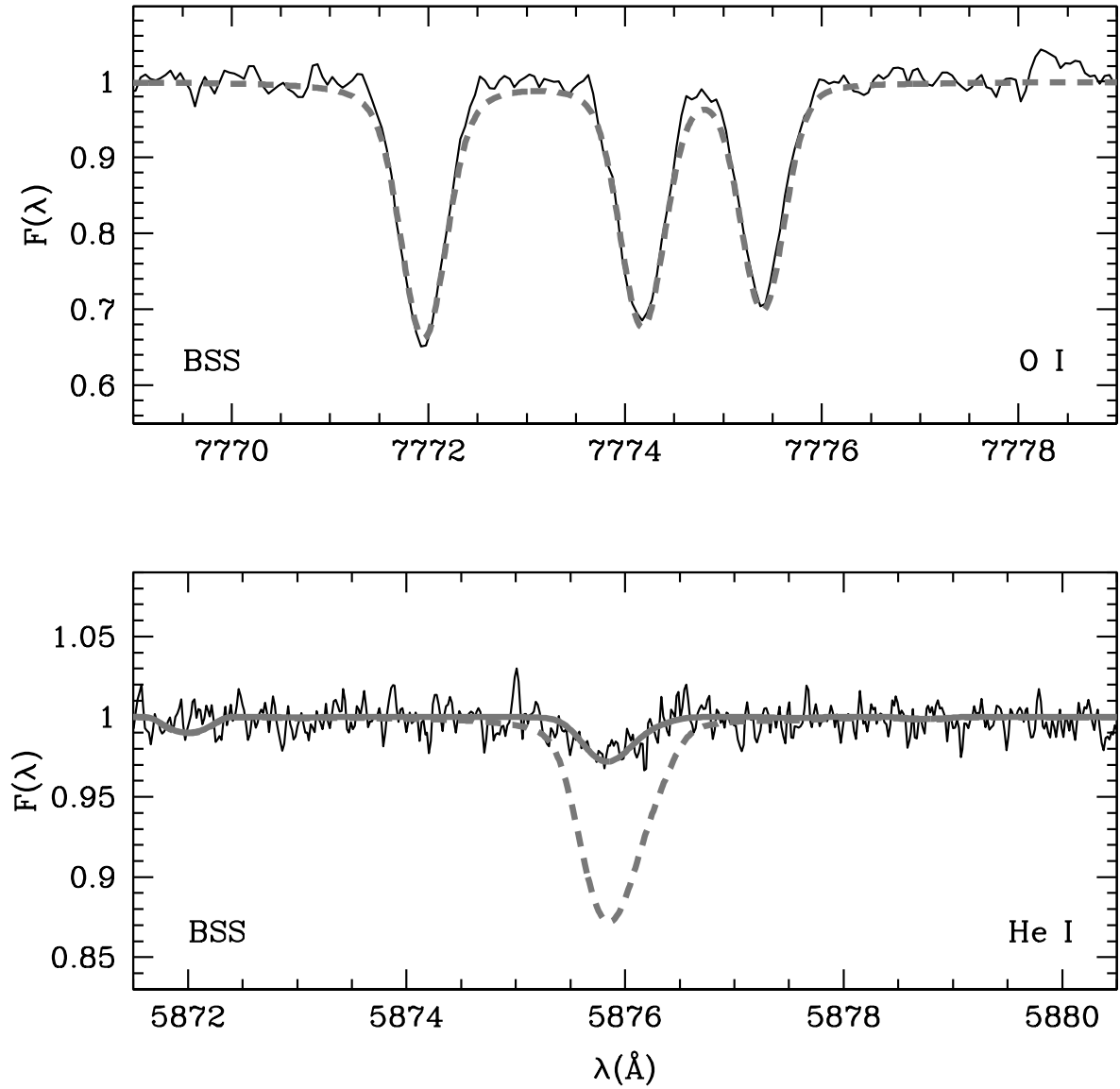}
\caption{Top panel: comparison between the observed spectrum of the BSS \#79976 (black line) and a synthetic spectrum with [O/H] = -0.44 dex (grey dashed line).
Bottom panel: He~I line at 5876 \AA\ for BSS \#1100063 compared with synthetic spectra with Y=0.25 (grey dashed line) and Y=0.001 (solid grey line).}
\label{oxrich}
\end{figure}

\begin{figure}
\plotone{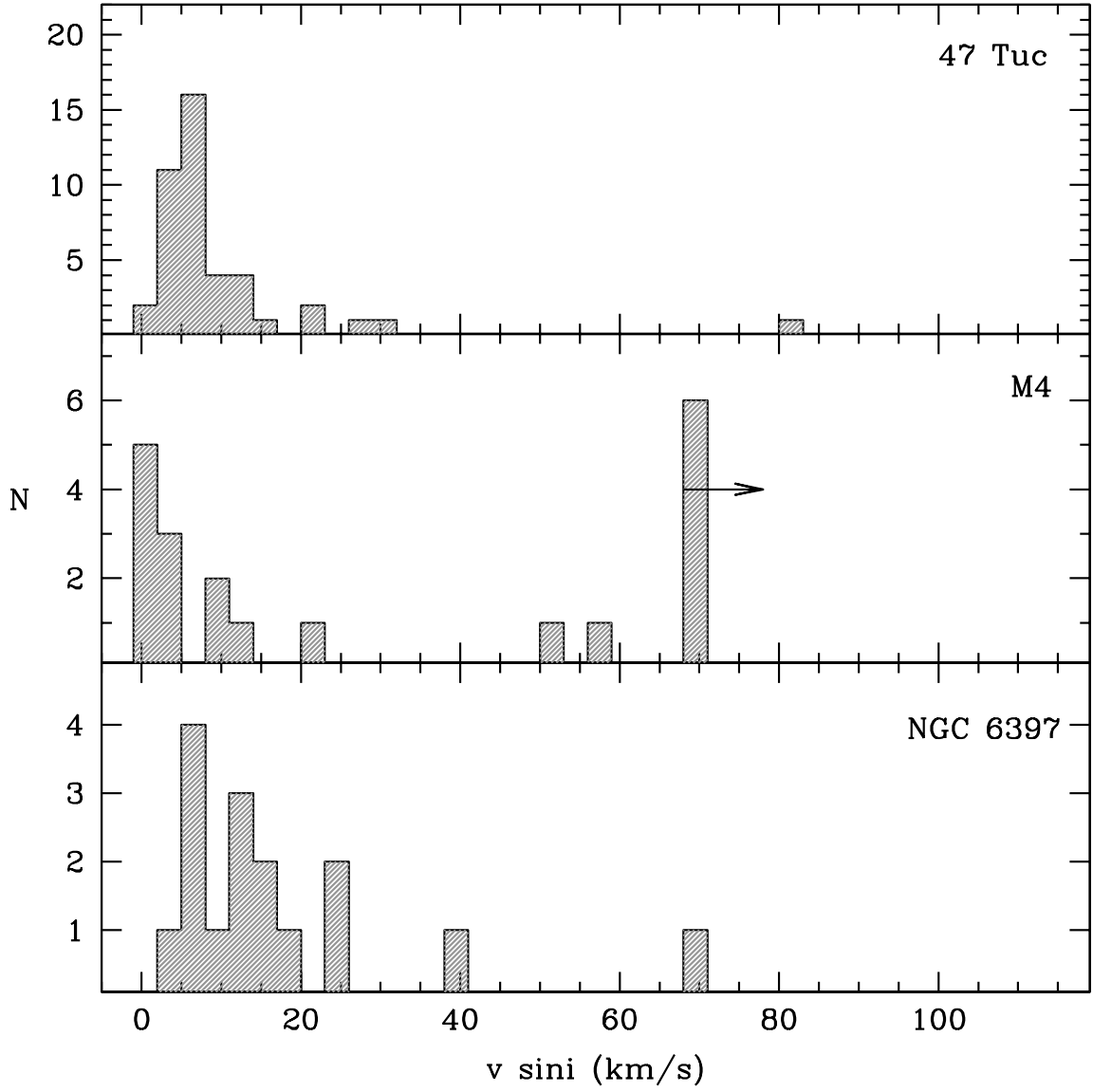}
\caption{Rotational velocity distributions for the BSS populations in 47 Tuc, M4 and NGC 6397. The arrow in the central panel indicates the lower limit
found for the fast rotators in M4.}
\label{rotconf}
\end{figure}

As for the metallicity, with respect to the TO sample, the hot
populations show a larger dispersion and a trend with the effective
temperature (see Figs. \ref{levconfgrey}, \ref{mgabb}, \ref{oxabb}).  
Such features have been already observed
in several GCs for HB stars hotter than $\sim 11000$ K
(\citealt{Behr1999, Behr2000a, Behr2003, Moehler2000, Fabbian2005,
  Pace2006, Hubrig2009}) and they are explained in terms of particle
transport mechanisms (as radiative levitation and gravitational
settling) occurring in the non-convective atmospheres of these stars
\citep[see, e.g.,][]{Richard2002, Michaud2008}. 
Metal enhancements with respect to the initial
composition are also observed in Population I main sequence stars
hotter than $\sim 7000-8000$ K, where the surface convective zone
starts to disappear \citep[see][and references therein]{Vick2010}.
Although BSSs in NGC 6397 belong to a different stellar population (they
are metal-poor, A-F type stars), a particle transport mechanism
occurring in absence of convection could also explain the
observational evidences presented here. 
% Indeed, the
% observed behaviour of [Fe/H] and [Mg/H] as a function of T
% (with increasing abundances in the hotter stars) indicates that
% element transport mechanisms driven by radiative levitation occur at a
% threshold temperature of $\sim 8200$ K. BSSs cooler than this
% temperature show photospheric abundances that well agree with the
% pristine composition of the cluster (as derived from the TO
% population), while for hotter stars significant metal enhancements are
% detected.  
Indeed, the
observed behaviour of [Fe/H] and [Mg/H] as a function of T
(with increasing abundances in the hotter stars) suggests that
element transport mechanisms driven by radiative levitation occur at a
threshold temperature of $\sim 8200$ K. Most of the BSSs cooler than this
temperature shows Fe and Mg abundances similar to the
pristine composition of the cluster (as derived from the TO
population), while for hotter stars significant metal enhancements are
detected.
Such a scenario is strengthened by the significant He
depletion observed in the hottest BSS of our sample. In fact, because
of the gravitational settling, He is progessively diffused downward in
the stellar interior, with a consequent reduction of its content on the
surface. Since the convection is sustained by the He opacity, the
convective zone starts to disappear and the elements with radiative
acceleration larger than the gravitational one are diffused upward and
enrich the photosphere with metals. Such a mechanism is particularly
efficient in absence of stellar rotation, that otherwise would inhibit
the metal levitation. Indeed, BSS \#1100063 displays the highest
temperature, a low rotational velocity and significant He depletion,
all concurring to produce its remarkably high Fe abundance
([Fe/H]=0.10).  Hence, while no theoretical models of levitation 
are currently available for stars similar to those
analysed here (i.e. metal-poor, A-F type stars), these results and the
fact that the onset of chemical anomalies in the BSS population occurs
at temperatures similar to those predicted for solar abundance MS stars, suggest
that BSSs behave like normal MS stars of the same spectral type.
However, for the hot BSSs we also measured a
significant enhancement of [Mg/H], that is not observed in HB and
Population I MS stars, likely because of a balance between radiative
and gravity accelerations.

While the observed chemical anomalies prevent us from drawing
conclusions about the BSS formation mechanisms in this cluster, the
present work provided us with the first information about the (still
poorly understood) particle transport mechanisms in a range of
metallicity and stellar mass not covered by other stellar systems.  In
this respect, similar studies of BSSs in metal-poor GCs, where the
temperature of these stars are high enough for radiative levitation
processes to occur, are highly desirable.

\acknowledgements
{This research is part of the project COSMIC-LAB (www.cosmic-lab.eu) 
funded by the European Research Council
(under contract ERC-2010-AdG-267675) and it was partly supported by PRIN INAF 2009 
``Formation and evolution of massive star clusters.''
We thank the anonymous referee for his/her valuable suggestions.}

\newpage
\begin{table}[H]
\centering
\begin{tabular}{c|c|c|c|c}
\hline
\hline
period & grating & element & N$_{exp}$ & time (h)\\
\hline
\hline
P73 & HR5A & Fe, Mg & 6 & 4.5 \\
    & HR18 & O      & 3 & 2   \\
    & UVES Red 580 & Fe, Mg & 9 & 6.5 \\
\hline
P81 & HR15N & H$\alpha$ & 2 & 1.5\\
    & HR18  & O     & 3 & 2  \\
\hline
P75 & HR5B  & Fe, Mg & 8 & 8 \\
\hline
\hline
\end{tabular}
\caption{Details about FLAMES observations for each observing run.
The adopted gratings, the sampled elements, the number of exposures and the total integration time have been listed.}
\label{gratings} 
\end{table}

\begin{sidewaystable}[H]
\scriptsize
\centering
\begin{tabular}{rccccccccccccl}
\hline
ID(P81) & RA & DEC & V & I & T & $\log(g)$ & M & RV & $v \sin i~~~~$ & [Fe/H] & [Mg/H] & [O/H]& Notes\\ 
        &  (degrees)  &  (degrees)   &   &   &(K)&           & (M$_{\odot}$) & ($\kms$)  & ($\kms$)   &&&\\
\hline

18705	& 264.9180785 & -53.5153682 & 15.55 & 15.16 & 8770 & 4.5 & 1.2 & $16.8\pm 0.7$ & $11\pm 1$& $-1.36\pm0.17$ & $-1.99\pm0.04$ &   --           &\\ 
62594	& 265.2697513 & -53.7701740 & 16.33 & 15.74 & 7568 & 4.5 & 1.0 & $17.6\pm 0.8$ & $ 6\pm 1$&   --           &   --           & $-1.88\pm0.07$ &\\
64782   & 265.1325939 & -53.7399209 & 14.99 & 14.55 & 8241 & 4.2 & 1.1 & $12.4\pm 0.4$ & $12\pm 1$& $-1.91\pm0.09$ & $-2.11\pm0.04$ & $-2.42\pm0.09$ &\\ 
75241	& 265.2277155 & -53.6491197 & 15.95 & 15.37 & 7727 & 4.4 & 1.0 & $13.0\pm 0.7$ & $ 5\pm 1$&   --           &   --           & $-1.97\pm0.14$ &\\ 
76278	& 265.2728458 & -53.6400582 & 15.47 & 15.12 & 9057 & 4.5 & 1.2 & $15.5\pm 0.6$ & $40\pm 1$& $-1.23\pm0.16$ & $-1.31\pm0.04$ & $-2.33\pm0.09$ &\\
79976	& 265.2213165 & -53.5989412 & 15.37 & 15.06 & 9311 & 4.5 & 1.3 & $24.0\pm 0.2$ & $ 7\pm 2$& $-1.15\pm0.07$ & $-1.31\pm0.05$ & $-0.44\pm0.06$ & O-rich\\ 
81828   & 265.2286114 & -53.5719809 & 15.19 & 14.76 & 8299 & 4.2 & 1.0 & $22.5\pm 0.4$ & $ 6\pm 3$& $ 0.09\pm0.06$ & $-2.56\pm0.05$ &   --           &\\
1100063 & 265.1757126 & -53.6745697 & 14.54 & 14.35 & 13183& 4.7 & 2.0 & $17.8\pm 0.1$ & $13\pm 6$& $ 0.10\pm0.06$ & $-0.22\pm0.18$ &   --           &\\ 
1100126	& 265.1891195 & -53.6735768 & 15.46 & 14.99 & 8204 & 4.3 & 1.0 & $30.8\pm 3.4$ & $70\pm 1$&   --           & $-1.94\pm0.04$ & $-2.24\pm0.02$ & FR, SX Phe\\
1100127 & 265.1837773 & -53.6778963 & 15.50 & 14.93 & 7551 & 4.1 & 0.9 & $23.8\pm 0.4$ & $15\pm 1$& $-1.98\pm0.06$ & $-1.87\pm0.05$ & $-1.91\pm0.06$ &SX Phe\\ 
1100157 & 265.1564245 & -53.6929579 & 15.66 & 15.03 & 7430 & 4.2 & 0.9 & $21.6\pm 0.4$ & $ 6\pm 1$&   --           &   --           & $-2.01\pm0.13$ &\\ 
1100162 & 265.2100723 & -53.6924561 & 15.62 & 15.22 & 8531 & 4.5 & 1.1 & $22.2\pm 0.5$ & $19\pm 1$& $-1.26\pm0.12$ & $-1.82\pm0.05$ & $-2.07\pm0.13$ &\\ 
1100170 & 265.1638009 & -53.6796128 & 15.68 & 15.18 & 7998 & 4.4 & 1.0 & $23.1\pm 0.5$ & $17\pm 1$&   --           &   --           & $-2.09\pm0.13$ &SX Phe\\ 
1100191 & 265.1816976 & -53.6743268 & 15.78 & 15.19 & 7638 & 4.3 & 0.9 & $15.9\pm 1.8$ &  --      &   --           &   --           &   --           &\\ 
1100208	& 265.1940391 & -53.6693609 & 15.78 & 15.39 & 8570 & 4.6 & 1.2 & $26.6\pm 0.3$ & $13\pm 1$& $-0.85\pm0.11$ & $-1.01\pm0.05$ & $-2.54\pm0.10$ &\\
2200110	& 265.1775032 & -53.6461613 & 15.18 & 14.70 & 8072 & 4.2 & 1.0 & $19.7\pm 0.3$ & $26\pm 1$& $-1.68\pm0.09$ & $-1.67\pm0.05$ & $-2.30\pm0.12$ &\\ 
2200239 & 265.1563541 & -53.6767039 & 15.97 & 15.50 & 8318 & 4.6 & 1.2 & $19.8\pm 0.8$ & $25\pm 4$& $-1.97\pm0.13$ & $-2.18\pm0.03$ & $-1.72\pm0.08$ &SX Phe\\ 
2200356 & 265.1635196 & -53.6463086 & 16.28 & 15.76 & 7816 & 4.6 & 1.1 & $23.4\pm 4.0$ &  --      &   --           &   --           &   --           &\\ 

\hline
\end{tabular}
 \caption{Coordinates, magnitudes, atmospherical parameters, masses, radial and rotational velocities, Fe, Mg and O abundances of the BSS sample.}
\label{bss} 
\end{sidewaystable}

\begin{sidewaystable}[H]
\scriptsize
\centering
\begin{tabular}{rcccccccccccc}
\hline
ID(P81) & RA & DEC & V & I & T & $\log(g)$ & M & RV & $v \sin i~~~~$ & [Fe/H] & [Mg/H] & [O/H]\\ 
        &  (degrees)  &  (degrees)   &   &   &(K)&           & (M$_{\odot}$) & ($\kms$)  & ($\kms$)   &&\\
\hline

474	  &  265.3579826 & -53.5451451 & 13.48 & 13.24  &   9978 & 3.6 & 0.6 & $19.2\pm  1.0$ & $38\pm1$ &   --           &	$-1.59\pm  0.04$ &$-1.74\pm  0.10$\\
1655	  &  265.3920971 & -53.5002154 & 13.14 & 12.80  &   8663 & 3.3 & 0.7 & $14.1\pm  0.4$ & $21\pm1$ &$ -2.26\pm 0.07$&	$-1.82\pm  0.09$ &$-2.10\pm  0.17$\\
9731	  &  265.2902124 & -53.5461587 & 13.36 & 13.06  &   9529 & 3.5 & 0.7 & $19.9\pm  0.3$ & $38\pm1$ &$ -1.97\pm 0.05$&	$-1.62\pm  0.05$ &$-1.38\pm  0.10$\\
10485	  &  265.1828792 & -53.5167585 & 13.72 & 13.49  &  10789 & 3.7 & 0.6 & $25.6\pm  0.9$ & $32\pm1$ &   --           &	$-1.81\pm  0.03$ &$-1.98\pm  0.06$\\
11063	  &  265.2707910 & -53.4935456 & 13.49 & 13.20  &   9952 & 3.6 & 0.6 & $18.2\pm  0.4$ & $33\pm1$ &   --           &	$-1.63\pm  0.04$ &$-1.42\pm  0.10$\\
11363	  &  265.1442501 & -53.4802352 & 13.73 & 13.51  &  10789 & 3.7 & 0.6 & $18.5\pm  0.5$ & $23\pm1$ &   --           &	$-1.86\pm  0.03$ &$-2.22\pm  0.05$\\
12158	  &  265.1336942 & -53.4423799 & 13.72 & 13.52  &  10757 & 3.7 & 0.6 & $15.5\pm  0.5$ & $ 9\pm1$ &$ -0.19\pm 0.14$&	$-2.00\pm  0.03$ &  --            \\
18540	  &  265.0135736 & -53.5217368 & 13.23 & 12.93  &   9050 & 3.4 & 0.7 & $19.5\pm  0.5$ & $40\pm1$ &   --           &	$-1.82\pm  0.04$ &$-1.80\pm  0.10$\\
51874	  &  264.9632133 & -53.6890207 & 13.77 & 13.56  &  10989 & 3.8 & 0.6 & $14.6\pm  0.2$ & $12\pm2$ &$  0.02\pm 0.16$&	$-1.77\pm  0.04$ &  --            \\
53805	  &  264.9638040 & -53.6514855 & 13.30 & 13.00  &   9299 & 3.4 & 0.7 & $22.6\pm  0.9$ & $36\pm1$ &   --           &	$-1.85\pm  0.03$ &$-1.69\pm  0.10$\\
54043	  &  265.0640413 & -53.6464338 & 13.64 & 13.37  &  10501 & 3.7 & 0.6 & $28.5\pm  0.9$ & $38\pm2$ &   --           &	$-1.78\pm  0.03$ &$-1.72\pm  0.08$\\
56322	  &  264.9978322 & -53.5980665 & 13.44 & 13.15  &   9805 & 3.5 & 0.6 & $21.7\pm  0.4$ & $41\pm1$ &   --           &	$-1.52\pm  0.05$ &$-1.27\pm  0.10$\\
57042	  &  265.0286135 & -53.5790632 & 13.55 & 13.28  &  10198 & 3.6 & 0.6 & $20.3\pm  0.9$ & $42\pm1$ &   --           &	$-1.50\pm  0.05$ &$-1.16\pm  0.10$\\
60940	  &  265.2943802 & -53.7943407 & 13.14 & 12.88  &   8787 & 3.3 & 0.7 & $24.0\pm  0.9$ & $38\pm1$ &$ -2.19\pm 0.07$&	$-1.84\pm  0.01$ &$-1.85\pm  0.11$\\
61653	  &  265.1776640 & -53.7844834 & 13.55 & 13.34  &  10227 & 3.6 & 0.6 & $28.4\pm  0.3$ & $31\pm1$ &   --           &	$-1.79\pm  0.04$ &$-1.97\pm  0.07$\\
63769	  &  265.1643436 & -53.7537368 & 13.10 & 12.80  &   8578 & 3.3 & 0.7 & $20.6\pm  0.1$ & $14\pm1$ &$ -2.34\pm 0.07$&	$-1.84\pm  0.04$ &$-2.24\pm  0.14$\\
64937	  &  265.2327151 & -53.7377808 & 13.43 & 13.17  &   9782 & 3.5 & 0.6 & $15.7\pm  0.4$ & $41\pm1$ &   --           &	$-1.67\pm  0.05$ &$-1.52\pm  0.10$\\
65798	  &  265.1531778 & -53.7272727 & 13.50 & 13.24  &  10004 & 3.6 & 0.6 & $22.8\pm  0.4$ & $20\pm1$ &   --           &	$-1.79\pm  0.04$ &$-2.18\pm  0.14$\\
70707	  &  265.2944846 & -53.6828826 & 13.33 & 13.07  &   9418 & 3.5 & 0.7 & $22.0\pm  0.3$ & $34\pm1$ &$ -2.26\pm 0.05$&	$-1.70\pm  0.04$ &$-1.59\pm  0.10$\\
72734	  &  265.2751355 & -53.6692360 & 13.43 & 13.19  &   9782 & 3.5 & 0.6 & $13.8\pm  0.2$ & $13\pm1$ &$ -2.09\pm 0.05$&	$-1.85\pm  0.04$ &$-2.37\pm  0.10$\\
73084	  &  265.2395639 & -53.6667402 & 13.12 & 12.76  &   8533 & 3.3 & 0.7 & $19.9\pm  0.2$ & $27\pm1$ &$ -2.14\pm 0.07$&	$-1.78\pm  0.05$ &$-1.54\pm  0.12$\\
73278	  &  265.2576032 & -53.6653473 & 13.66 & 13.43  &  10596 & 3.7 & 0.6 & $30.0\pm  1.1$ & $38\pm1$ &   --           &	$-1.80\pm  0.03$ &$-2.14\pm  0.05$\\
74139	  &  265.3052140 & -53.6585939 & 13.58 & 13.36  &  10316 & 3.6 & 0.6 & $22.8\pm  0.9$ & $42\pm1$ &   --           &	$-1.75\pm  0.04$ &$-1.88\pm  0.08$\\
74430	  &  265.2862434 & -53.6562185 & 13.02 & 12.64  &   8151 & 3.2 & 0.7 & $19.8\pm  0.2$ & $26\pm1$ &$ -2.20\pm 0.08$&	$-1.78\pm  0.05$ &$-1.45\pm  0.13$\\
74883	  &  265.2546642 & -53.6522206 & 13.47 & 13.21  &   9926 & 3.6 & 0.6 & $20.0\pm  0.2$ & $16\pm1$ &$ -2.25\pm 0.06$&	$-1.84\pm  0.04$ &$-2.05\pm  0.13$\\
76169 	  &  265.3280914 & -53.6409329 & 13.67 & 13.45  &  10628 & 3.7 & 0.6 & $27.0\pm  1.1$ & $40\pm2$ &   --           &	$-1.80\pm  0.04$ &$-2.04\pm  0.06$\\
76330	  &  265.1989537 & -53.6396150 & 13.53 & 13.29  &  10141 & 3.6 & 0.6 & $30.6\pm  0.4$ & $35\pm1$ &   --           &	$-1.70\pm  0.03$ &$-1.66\pm  0.10$\\
76738	  &  265.1755658 & -53.6357717 & 13.61 & 13.37  &  10407 & 3.7 & 0.6 & $31.6\pm  1.0$ & $36\pm1$ &   --           &	$-1.78\pm  0.03$ &$-1.86\pm  0.08$\\
77082	  &  265.2399386 & -53.6322805 & 13.57 & 13.33  &  10286 & 3.6 & 0.6 & $14.0\pm  0.6$ & $40\pm1$ &   --           &	$-1.81\pm  0.03$ &$-1.63\pm  0.10$\\
77542	  &  265.1136890 & -53.6274217 & 13.31 & 13.02  &   9349 & 3.4 & 0.6 & $13.2\pm  0.1$ & $10\pm2$ &$ -2.12\pm 0.04$&	$-1.86\pm  0.04$ &$-2.32\pm  0.13$\\
77637	  &  265.2727433 & -53.6263464 & 13.09 & 12.76  &   8506 & 3.3 & 0.7 & $19.6\pm  0.7$ & $37\pm1$ &$ -2.25\pm 0.12$&	$-1.80\pm  0.05$ &$-1.72\pm  0.11$\\
80792	  &  265.1226697 & -53.5878451 & 13.41 & 13.16  &   9737 & 3.5 & 0.7 & $14.7\pm  0.3$ & $17\pm1$ &$ -2.14\pm 0.05$&	$-1.84\pm  0.04$ &$-2.36\pm  0.06$\\
81455	  &  265.1657796 & -53.5774433 & 13.77 & 13.55  &  10989 & 3.8 & 0.6 & $13.3\pm  0.2$ & $15\pm1$ &$ -1.93\pm 0.08$&	$-1.84\pm  0.03$ &$-2.33\pm  0.08$\\
82001	  &  265.1649280 & -53.5693729 & 13.20 & 12.89  &   8911 & 3.4 & 0.7 & $18.2\pm  0.2$ & $ 9\pm1$ &$ -2.16\pm 0.07$&	$-1.86\pm  0.04$ &$-2.22\pm  0.13$\\
82561	  &  265.2503688 & -53.5602131 & 13.56 & 13.31  &  10227 & 3.6 & 0.6 & $23.9\pm  1.0$ & $35\pm1$ &   --           &	$-1.69\pm  0.04$ &$-1.82\pm  0.08$\\
89014	  &  265.4087478 & -53.7085078 & 13.15 & 12.87  &   8820 & 3.3 & 0.7 & $25.2\pm  0.1$ & $ 9\pm2$ &$ -2.23\pm 0.05$&	$-1.85\pm  0.04$ &$-2.20\pm  0.17$\\
90302	  &  265.4928341 & -53.6737610 & 13.43 & 13.18  &   9759 & 3.5 & 0.6 & $22.5\pm  0.4$ & $28\pm1$ &$ -2.13\pm 0.11$&	$-1.73\pm  0.04$ &$-1.64\pm  0.09$\\
91315	  &  265.4124151 & -53.6457029 & 13.52 & 13.29  &  10085 & 3.6 & 0.6 & $22.4\pm  0.8$ & $41\pm1$ &   --           &	$-1.75\pm  0.04$ &$-1.81\pm  0.09$\\
1100024   &  265.1680268 & -53.6816479 & 13.54 & 13.37  &  10198 & 3.6 & 0.6 & $23.7\pm  0.2$ & $ 7\pm2$ &$ -2.05\pm 0.04$&	$-1.87\pm  0.03$ &$-2.37\pm  0.09$\\
1100029   &  265.1822376 & -53.6849407 & 13.64 & 13.43  &  10501 & 3.7 & 0.6 & $19.5\pm  0.2$ & $ 8\pm1$ &$ -1.91\pm 0.05$&	$-1.83\pm  0.04$ &$-2.38\pm  0.07$\\
2200024   &  265.1991228 & -53.6643097 & 13.43 & 13.20  &   9782 & 3.5 & 0.6 & $21.9\pm  0.3$ & $23\pm1$ &$ -1.96\pm 0.05$&	$-1.64\pm  0.04$ &$-1.90\pm  0.16$\\
2200028   &  265.1423624 &  -53.6557501& 13.54 & 13.29  &  10169 & 3.6 & 0.6 & $18.9\pm  0.4$ & $38\pm1$ &   --           &	$-1.74\pm  0.03$ &$-1.92\pm  0.08$\\
\hline
\end{tabular}
 \caption{Coordinates, magnitudes, atmospherical parameters, masses, radial and rotational velocities, Fe, Mg and O abundances of the HB sample.}
\label{hb} 
\end{sidewaystable}

\begin{sidewaystable}[H]
\scriptsize
\centering
\begin{tabular}{rccccccccccc}
\hline
ID(P81) & RA & DEC & V & I & T & $\log(g)$ & M & RV & $v \sin i~~~~$ & [Fe/H] & [Mg/H]\\ 
        &  (degrees)  &  (degrees)   &   &   &(K)&           & (M$_{\odot}$) & ($\kms$)  & ($\kms$)   &&\\
\hline
 9407 & 265.1440384 & -53.5557399 & 15.56 & 14.54 & 5546 & 3.5 & 0.7 & $17.2\pm0.10$ & $7\pm4$& $-2.10\pm0.20$  & $-1.93\pm0.10$\\
17909 & 264.9203310 & -53.5488131 & 16.26 & 15.43 & 6546 & 4.1 & 0.7 & $21.9\pm0.10$ & $6\pm3$& $-2.21\pm0.22$  & $-1.82\pm0.04$\\
41395 & 264.8446941 & -53.6477354 & 15.79 & 14.76 & 5728 & 3.6 & 0.7 & $18.4\pm0.12$ & $4\pm3$& $-2.06\pm0.20$  & $-1.82\pm0.10$\\
41662 & 264.8421933 & -53.6398126 & 15.83 & 14.85 & 5848 & 3.7 & 0.7 & $19.2\pm0.12$ & $7\pm4$& $-2.17\pm0.20$  & $-1.85\pm0.10$\\
43151 & 264.8440820 & -53.5979440 & 16.05 & 15.13 & 6281 & 3.9 & 0.7 & $20.3\pm0.12$ & $7\pm3$& $-2.21\pm0.21$  & $-1.75\pm0.06$\\
46551 & 265.0716357 & -53.8067644 & 16.22 & 15.35 & 6501 & 4.0 & 0.7 & $18.7\pm0.19$ & $7\pm3$& $-2.15\pm0.22$  & $-2.00\pm0.04$\\
48483 & 265.0570850 & -53.7600905 & 15.77 & 14.78 & 5728 & 3.6 & 0.7 & $19.2\pm0.11$ & $8\pm4$& $-2.19\pm0.20$  & $  -- \pm0.10$\\
48487 & 265.0266095 & -53.7599815 & 16.00 & 15.09 & 6237 & 3.9 & 0.7 & $24.8\pm0.11$ & $7\pm3$& $-2.22\pm0.21$  & $-1.93\pm0.06$\\
48562 & 265.0198484 & -53.7584877 & 15.96 & 15.03 & 6138 & 3.8 & 0.7 & $19.9\pm0.11$ & $6\pm3$& $-2.18\pm0.21$  & $-1.86\pm0.06$\\
48646 & 265.0818209 & -53.7564627 & 15.95 & 15.05 & 6138 & 3.8 & 0.7 & $23.5\pm0.10$ & $6\pm3$& $-2.21\pm0.21$  & $-1.91\pm0.06$\\
\hline
\end{tabular}
 \caption{Coordinates, magnitudes, atmospherical parameters, masses, radial and rotational velocities, Fe and Mg abundances of the TO sample.
A complete version of the table is available in electronic form.}
\label{to} 
\end{sidewaystable}


\newpage
\begin{thebibliography}{}
\bibitem[Behr et al.(1999)]{Behr1999} Behr, B.~B., Cohen, J.~G., McCarthy, J.~K., \& Djorgovski, S.~G.\ 1999, \apjl, 517, L135 
\bibitem[Behr et al.(2000a)]{Behr2000a} Behr, B.~B., Cohen, J.~G., \& McCarthy, J.~K.\ 2000a, \apjl, 531, L37 
\bibitem[Behr et al.(2000b)]{Behr2000b} Behr, B.~B., Djorgovski, S.~G., Cohen, J.~G., et al.\ 2000b, \apj, 528, 849 
\bibitem[Behr(2003)]{Behr2003} Behr, B.~B.\ 2003, \apjs, 149, 67 
\bibitem[Benz \& Hills(1987)]{BenzHills1987} Benz, W., \& Hills, J.~G.\ 1987, \apj, 323, 614 
\bibitem[Caffau et al.(2011)]{Caffau2011} Caffau, E., Ludwig, H.-G., Steffen, M., Freytag, B., \& Bonifacio, P.\ 2011, \solphys, 268, 255 
\bibitem[Carretta et al.(2009a)]{Carretta2009a} Carretta, E., Bragaglia, A., Gratton, R., \& Lucatello, S.\ 2009a, \aap, 505, 139 
\bibitem[Carretta et al.(2009b)]{Carretta2009b} Carretta, E., Bragaglia, A., Gratton, R.~G., et al.\ 2009b, \aap, 505, 117 
\bibitem[Carretta et al.(2010a)]{Carretta2010a} Carretta, E., Bragaglia, A., Gratton, R.~G., et al.\ 2010a, \aap, 520, A95 
\bibitem[Carretta et al.(2010b)]{Carretta2010b} Carretta, E., Gratton, R.~G., Lucatello, S., et al.\ 2010b, \apjl, 722, L1 
\bibitem[Carretta et al.(2010c)]{Carretta2010c} Carretta, E., Bragaglia, A., Gratton, R., et al.\ 2010c, \apjl, 712, L21 
\bibitem[Carretta et al.(2011)]{Carretta2011} Carretta, E., Lucatello, S., Gratton, R.~G., Bragaglia, A., \& D'Orazi, V.\ 2011, \aap, 533, A69 
\bibitem[Castelli et al.(1997)]{Castelli1997} Castelli, F., Gratton, R.~G., \& Kurucz, R.~L.\ 1997, \aap, 318, 841 
\bibitem[Castelli \& Kurucz(2003)]{CastelliKurucz2003} Castelli, F., \& Kurucz, R.~L.\ 2003, Modelling of Stellar Atmospheres, 210, 20P 
\bibitem[Castelli(2005)]{Castelli2005} Castelli, F.\ 2005, Memorie della Societa Astronomica Italiana Supplementi, 8, 25 
\bibitem[Charbonneau \& Michaud(1991)]{Charbonneau1991} Charbonneau, P., \& Michaud, G.\ 1991, \apj, 370, 693 
\bibitem[Cohen \& McCarthy(1997)]{Cohen97} Cohen, J.~G., \& McCarthy, J.~K.\ 1997, \aj, 113, 1353 
\bibitem[D'Antona \& Caloi(2004)]{Dantona2004} D'Antona, F., \& Caloi, V.\ 2004, \apj, 611, 871 
\bibitem[De Marco et al.(2005)]{DeMarco2005} De Marco, O., Shara, M.~M., Zurek, D., Ouellette, J.~A., Lanz, T., Saffer, R.~A., \& Sepinsky, J.~F.\ 2005, \apj, 632, 894 
\bibitem[Fabbian et  al.(2005)]{Fabbian2005} Fabbian, D., Recio-Blanco, A., Gratton, R.~G., \& Piotto, G.\ 2005, \aap, 434, 235 
\bibitem[Ferraro et al.(1999)]{Ferraro1999} Ferraro, F.~R., Messineo, M., Fusi Pecci, F., et al.\ 1999, \aj, 118, 1738 
\bibitem[Ferraro et al.(2006)]{Ferraro2006} Ferraro, F.~R., Sabbi, E., Gratton, R., et al.\ 2006, \apjl, 647, L53 (F06)
\bibitem[Ferraro et al.(2009a)]{Ferraro2009a} Ferraro, F.~R., Dalessandro, E., Mucciarelli, A., et al.\ 2009a, \nat, 462, 483 
\bibitem[Ferraro et al.(2009b)]{Ferraro2009b} Ferraro, F.~R., Beccari, G., Dalessandro, E., et al.\ 2009b, \nat, 462, 1028 
\bibitem[For \& Sneden(2010)]{ForSneden2010} For, B.-Q., \& Sneden, C.\ 2010, \aj, 140, 1694 
\bibitem[Gebran \& Monier(2008)]{GebranMonier2008} Gebran, M., \& Monier, R.\ 2008, \aap, 483, 567 
\bibitem[Gebran et al.(2008)]{Gebran2008} Gebran, M., Monier, R., \& Richard, O.\ 2008, Contributions of the Astronomical Observatory Skalnate Pleso, 38, 405 
\bibitem[Gebran et al.(2010)]{Gebran2010} Gebran, M., Vick, M., Monier, R., \& Fossati, L.\ 2010, \aap, 523, A71 
\bibitem[Glaspey et al.(1986)]{Glaspey86} Glaspey, J.~W., Demers, S., Moffat, A.~F.~J., \& Michaud, G.\ 1986, \pasp, 98, 1123 
\bibitem[Glaspey et al.(1989)]{Glaspey89} Glaspey, J.~W., Michaud, G., Moffat, A.~F.~J., \& Demers, S.\ 1989, \apj, 339, 926 
\bibitem[Gonz{\'a}lez Hern{\'a}ndez et al.(2010)]{GonzalezHernandez2010} Gonz{\'a}lez Hern{\'a}ndez, J.~I., et al.\ 2010, IAU Symposium, 268, 257 
\bibitem[Gratton et al.(1999)]{Gratton1999} Gratton, R.~G., Carretta, E., Eriksson, K., \& Gustafsson, B.\ 1999, \aap, 350, 955 
\bibitem[Gratton et al.(2001)]{Gratton2001} Gratton, R.~G., Bonifacio, P., Bragaglia, A., et al.\ 2001, \aap, 369, 87 
\bibitem[Gratton et al.(2004)]{Gratton2004} Gratton, R., Sneden, C., \& Carretta, E.\ 2004, \araa, 42, 385 
\bibitem[Gratton et al.(2011)]{Gratton2011} Gratton, R.~G., Lucatello, S., Carretta, E., et al.\ 2011, \aap, 534, A123 
\bibitem[Gratton et al.(2012)]{Gratton2012} Gratton, R.~G., Lucatello, S., Carretta, E., et al.\ 2012, \aap, 539, A19 
\bibitem[Gray(1984)]{Gray1984} Gray, D.~F.\ 1984, \apj, 281, 719 
\bibitem[Grevesse \& Sauval(1998)]{GrevesseSauval1998} Grevesse, N., \& Sauval, A.~J.\ 1998, \ssr, 85, 161 
\bibitem[Grundahl et al.(1999)]{Grundhal1999} Grundahl, F., Catelan, M., Landsman, W.~B., Stetson, P.~B., \& Andersen, M.~I.\ 1999, \apj, 524, 242 \bibitem[Harris(1996)]{Harris1996} Harris, W.~E.\ 1996, \aj, 112, 1487 
\bibitem[Hills \& Day (1976)]{hills76_colbss} Hills, J.~G., \& Day,
  C.~A.\ 1976, \aplett, 17, 87
\bibitem[Hubrig et al.(2009)]{Hubrig2009} Hubrig, S., Castelli, F., de Silva, G., Gonz{\'a}lez, J.~F., Momany, Y., Netopil, M., \& Moehler, S.\ 2009, \aap, 499, 865 
\bibitem[Johnson \& Pilachowski(2010)]{Johnson2010} Johnson, C.~I., \& Pilachowski, C.~A.\ 2010, \apj, 722, 1373 
\bibitem[Kaluzny \& Thompson(2003)]{Kaluzny2003} Kaluzny, J., \& Thompson, I.~B.\ 2003, \aj, 125, 2534 
\bibitem[Kurucz(1993)]{Kurucz1993} Kurucz, R. L. 1993, CD-ROM 13, 18 http://kurucz.harvard.edu
\bibitem[Lambert et al.(1992)]{Lambert92} Lambert, D.~L., McWilliam, A., \& Smith, V.~V.\ 1992, \apj, 386, 685 
\bibitem[Leonard \& Livio(1995)]{LeonardLivio1995} Leonard, P.~J.~T., \& Livio, M.\ 1995, \apjl, 447, L121 
\bibitem[Lind et al.(2008)]{Lind2008} Lind, K., Korn, A.~J., Barklem, P.~S., \& Grundahl, F.\ 2008, \aap, 490, 777 
\bibitem[Lind et al.(2011)]{lind2011} Lind, K., Charbonnel, C., Decressin, T., Primas, F., Grundahl, F., \& Asplund, M., 2011, \aap, 527, 148
\bibitem[Lombardi et al.(1995)]{Lombardi1995} Lombardi, J., C., Jr., Rasio, F.~A., \& Shapiro, S.~L.\ 1995, \apjl, 445, L117 
\bibitem[Lovisi et al.(2010)]{Lovisi2010} Lovisi, L., Mucciarelli, A., Ferraro, F.~R., et al.\ 2010, \apjl, 719, L121 (L10)
\bibitem[Lucatello \& Gratton(2003)]{LucatelloGratton2003} Lucatello, S., \& Gratton, R.~G.\ 2003, \aap, 406, 691 
\bibitem[Marino et al.(2009)]{Marino2009} Marino, A.~F., Milone, A.~P., Piotto, G., et al.\ 2009, \aap, 505, 1099 
\bibitem[Marino et al.(2011a)]{Marino2011a} Marino, A.~F., Sneden, C., Kraft, R.~P., et al.\ 2011a, \aap, 532, A8 
\bibitem[Marino et al.(2011b)]{Marino2011b} Marino, A.~F., Villanova, S., Milone, A.~P., et al.\ 2011b, \apjl, 730, L16 
\bibitem[Mashonkina et al.(2011)]{Mashonkina2011} Mashonkina, L., Gehren, T., Shi, J.-R., Korn, A.~J., \& Grupp, F.\ 2011, \aap, 528, A87 
\bibitem[McCrea (1964)]{mcrea64_mtbss1} McCrea, W.~H.\ 1964, \mnras,
  128, 147
\bibitem[Milone et al.(2006)]{Milone2006} Milone, A.~P., Villanova, S., Bedin, L.~R., Piotto, G., Carraro, G., Anderson, J., King, I.~R., \& Zaggia, S.\ 2006, \aap, 456, 517 
\bibitem[Michaud et al.(1983)]{Michaud1983} Michaud, G., Vauclair, G., \& Vauclair, S.\ 1983, \apj, 267, 256
\bibitem[Michaud et al.(2008)]{Michaud2008} Michaud, G., Richer, J., \& Richard, O.\ 2008, \apj, 675, 1223 
\bibitem[Moehler et al.(2000)]{Moehler2000} Moehler, S., Sweigart, A.~V., Landsman, W.~B., \& Heber, U.\ 2000, \aap, 360, 120 
\bibitem[Monier(2005)]{Monier2005} Monier, R.\ 2005, \aap, 442, 563 
\bibitem[Mucciarelli et al.(2009)]{Mucciarelli2009} Mucciarelli, A., Origlia, L., Ferraro, F.~R., \& Pancino, E.\ 2009, \apjl, 695, L134 
\bibitem[Mucciarelli et al.(2011)]{Mucciarelli2011} Mucciarelli, A., Salaris, M., Lovisi, L., et al.\ 2011, \mnras, 412, 81 
\bibitem[Norris \& Da Costa(1995)]{NorrisDaCosta1995} Norris, J.~E., \& Da Costa, G.~S.\ 1995, \apj, 447, 680 
\bibitem[Origlia et al.(2011)]{Origlia2011} Origlia, L., Rich, R.~M., Ferraro, F.~R., et al.\ 2011, \apjl, 726, L20 
\bibitem[Osterbrock et al.(1996)]{Osterbrock1996} Osterbrock, D.~E., Fulbright, J.~P., Martel, A.~R., Keane, M.~J., Trager, S.~C., \& Basri, G.\ 1996, \pasp, 108, 277
\bibitem[Pace et al.(2006)]{Pace2006} Pace, G., Recio-Blanco, A., Piotto, G., \& Momany, Y.\ 2006, \aap, 452, 493 
\bibitem[Peterson et al.(1995)]{Peterson1995} Peterson, R.~C., Rood, R.~T., \& Crocker, D.~A.\ 1995, \apj, 453, 214 
\bibitem[Peterson et al.(2000)]{Peterson2000} Peterson, R.~C., Rood, R.~T., Crocker, D.~A., \& Kraft, R.~P.\ 2000, Liege International Astrophysical Colloquia, 35, 523 
\bibitem[Pietrinferni et al.(2006)]{Pietrinferni2006} Pietrinferni, A., Cassisi, S., Salaris, M., \& Castelli, F.\ 2006, \apj, 642, 797 
\bibitem[Przybilla et al.(2001)]{Przybilla2001} Przybilla, N., Butler, K., Becker, S.~R., \& Kudritzki, R.~P.\ 2001, \aap, 369, 1009 
\bibitem[Recio-Blanco et al.(2002)]{RecioBlanco2002} Recio-Blanco, A., Piotto, G., Aparicio, A., \& Renzini, A.\ 2002, \apjl, 572, L71 
\bibitem[Recio-Blanco et al.(2004)]{RecioBlanco2004} Recio-Blanco, A., Piotto, G., Aparicio, A., \& Renzini, A.\ 2004, \aap, 417, 597 
\bibitem[Renzini \& Buzzoni(1986)]{Renzini1986} Renzini, A., \& Buzzoni, A.\ 1986, Spectral Evolution of Galaxies, 122, 195 
\bibitem[Richard et al.(2002)]{Richard2002} Richard, O., Michaud, G., Richer, J., et al.\ 2002, \apj, 568, 979 
\bibitem[Sarna \& De Greve(1996)]{SarnaDeGreve1996} Sarna, M.~J., \& De Greve, J.-P.\ 1996, \qjras, 37, 11 
\bibitem[Sbordone et al.(2004)]{Sbordone2004} Sbordone, L., Bonifacio, P., Castelli, F., \& Kurucz, R.~L.\ 2004, Memorie della Societa Astronomica Italiana Supplementi, 5, 93 
\bibitem[Shara et al.(1997)]{Shara1997} Shara, M.~M., Saffer, R.~A., \& Livio, M.\ 1997, \apjl, 489, L59 
\bibitem[Sills \& Pinsonneault(2000)]{SillsPinsonneault2000} Sills, A., \& Pinsonneault, M.~H.\ 2000, \apj, 540, 489 
\bibitem[Sills et al.(2005)]{Sills2005} Sills, A., Adams, T., \& Davies, M.~B.\ 2005, \mnras, 358, 716 
\bibitem[Takeda(1997)]{Takeda1997} Takeda, Y.\ 1997, \pasj, 49, 471
\bibitem[Th{\'e}venin et al.(2001)]{Thevenin2001} Th{\'e}venin, F., Charbonnel, C., de Freitas Pacheco, J.~A., et al.\ 2001, \aap, 373, 905 
\bibitem[Tonry \& Davis(1979)]{TonryDavis1979} Tonry, J., \& Davis, M.\ 1979, \aj, 84, 1511
\bibitem[Varenne \& Monier(1999)]{VarenneMonier1999} Varenne, O., \& Monier, R.\ 1999, \aap, 351, 247 
\bibitem[Vauclair(1999)]{Vauclair1999} Vauclair, S.\ 1999, \aap, 351, 973 
\bibitem[Vick et al.(2010)]{Vick2010} Vick, M., Michaud, G., Richer, J., \& Richard, O.\ 2010, \aap, 521, A62 
\bibitem[Vink \& Cassisi(2002)]{VinkCassisi} Vink, J.~S., \& Cassisi, S.\ 2002, \aap, 392, 553 
\bibitem[Zinn \& Searle (1976)]{zinn76_mtbss2} Zinn, R., \& Searle, L.\ 1976, \apj,
209, 734
\end{thebibliography}
\end{document}